\documentclass[fleqn,usenatbib]{mnras}

\usepackage{newtxtext,newtxmath}

\usepackage[T1]{fontenc}
\usepackage{ae,aecompl}

\usepackage{graphicx}	
\usepackage{amsmath}	
\usepackage{amssymb}	
\usepackage[dvipsnames]{xcolor}
\usepackage{xparse}

\usepackage{xspace}

\newcommand\newedit[1]{{#1}}

\newcommand\ba{\begin{eqnarray}}
\newcommand\ea{\end{eqnarray}}

\newcommand{\athena}{\texttt{Athena++}\xspace}
\newcommand{\M}{{\ensuremath{\mathcal{M}}}\xspace}
\newcommand{\CS}{{\ensuremath{C_{\rm S}}}\xspace}
\newcommand{\ave}[1]{\left<#1\right>}

\NewDocumentCommand\pder{mmg}{\ensuremath{
		\IfNoValueTF{#3}
		{\dfrac{\partial #1}{\partial #2}}
		{\left(\dfrac{\partial #1}{\partial #2}\right)_{#3}}
}}

\title[Wave-Driven Transport]{Boundary Layers of Accretion Discs: Wave-Driven Transport and Disc Evolution}

\author[Coleman et al.]{
Matthew S. B. Coleman,$^{1,2}$\thanks{E-mail: msbc@astro.princeton.edu}
Roman R. Rafikov,$^{1,3}$\thanks{John N. Bahcall Fellow at the IAS}
Alexander A. Philippov$^{4}$
\\
$^{1}$Institute for Advanced Study, Einstein Drive, Princeton, NJ 08540, USA\\
$^{2}$Department of Astrophysical Sciences, 4 Ivy Lane, Princeton University, Princeton, NJ 08540, USA\\
$^{3}$Centre for Mathematical Sciences, Department of Applied Mathematics and Theoretical Physics, University of Cambridge, \\ \hspace{.05em} Wilberforce Road, Cambridge CB3 0WA, UK\\
$^{4}$Center for Computational Astrophysics, Flatiron Institute, 162 Fifth Avenue, New York, NY 10010, USA
}

\date{Accepted XXX. Received YYY; in original form ZZZ}

\pubyear{2015}

\begin{document}
\label{firstpage}
\pagerange{\pageref{firstpage}--\pageref{lastpage}}
\maketitle

\defcitealias{Coleman2021}{Paper I}

\begin{abstract}
Astrophysical objects possessing a material surface (white dwarfs, \newedit{ young} stars, etc.) may accrete gas from the disc through the so-called surface boundary layer (BL), in which the angular velocity of the accreting gas experiences a sharp drop. Acoustic waves excited by the supersonic shear in the BL play an important role in mediating the angular momentum and mass transport through that region. Here we examine the characteristics of the angular momentum transport produced by the different types of wave modes emerging in the inner disc, using the results of a large suite of hydrodynamic simulations of the BLs. We provide a comparative analysis of the transport properties of different modes across the range of relevant disc parameters. In particular, we identify the types of modes which are responsible for the mass accretion onto the central object. {We find the correlated perturbations of surface density and radial velocity to provide an important contribution to the mass accretion rate. Although the wave-driven transport is intrinsically non-local, we do observe a clear correlation between the angular momentum flux injected into the disc by the waves and the mass accretion rate through the BL. We find the efficiency of angular momentum transport (normalized by thermal pressure) to be a weak function of the flow Mach number.} We also quantify the wave-driven evolution of the inner disc, in particular the modification of the angular frequency profile in the disc. Our results pave the way for understanding wave-mediated transport in future three-dimensional, magnetohydrodynamic studies of the BLs.
\end{abstract}

\begin{keywords}
accretion, accretion discs -- hydrodynamics -- instabilities
\end{keywords}



\section{Introduction}
\label{sect:intro}


Disc accretion onto an object possessing a material surface (i.e. not a black hole) is an important problem emerging in a variety of astrophysical settings. When the magnetic field of an accretor is weak enough \citep{Ghosh1978,Kon1991}, the accretion flow can extend all the way to its surface. In this case the transfer of incoming matter onto the central object (hereafter "a star") must proceed through the so-called {\it boundary layer} (hereafter BL) --- a narrow region between the disc and the accretor, in which the angular velocity of the accreting matter adjusts to the rotation of the central star. In order for that to happen, the gas arriving from the disc must lose its angular momentum inside the BL. 

The magneto-rotational instability (MRI; \citealt{VEL59,BAL91}), traditionally invoked as the angular momentum transport mechanism in sufficiently ionized accretion discs, cannot enable this process in the BL: it does not operate in this region \citep{PES12} since the angular velocity of the fluid there necessarily {\it increases} with the distance. Instead, \citet{BR12} proposed that a {\it sonic instability}, driving the excitation of acoustic waves by the supersonic shear flow inside the BL, mediates the angular momentum transport in that region in a {\it non-local} fashion, very different from the local $\alpha$-models \citet{SS73}, that have been previously invoked in the BL context \citep{POP93,BAL09,Hert2017}. This instability has been subsequently verified to robustly operate within the BL and drive angular momentum transport using hydrodynamic and magneto-hydrodynamic (MHD) simulations \citep{BRS12,BRS13a,BRS13b}. 

Recently \citet{Coleman2021} (hereafter \citetalias{Coleman2021}) presented a new, extensive suite of two-dimensional (2D) hydrodynamic simulations of the BLs run for multiple values of the Mach number $\M=v_K(R_\star)/c_s$ --- the ratio of the Keplerian velocity $v_K(R_\star)$ at the inner edge of the disc (i.e. at the stellar radius $R_\star$) to the sound speed $c_s$ (which was assumed constant in that work). They carefully analyzed the different waves emerging in the inner disc as a result of the sonic instability operating in the BL, and discovered some new types of modes emerging in the vicinity of the BL, e.g. the vortex-driven modes. That study focused predominantly on the {\it morphological} characteristics of the waves and their correspondence to the known analytical results.

In this work we will use the simulation suite presented in \citetalias{Coleman2021} to characterize the BL from a different angle, namely to explore the angular momentum and mass transport in its vicinity driven by the wave activity. A number of past studies explored the connection between the two transport processes in accretion discs. For example, \citet{BalPap1999} argued that the MRI-driven transport can be characterized as a local $\alpha$-viscosity \citep{SS73}, whereas the transport driven by disc self-gravity is global and cannot be represented using the $\alpha$-anzatz (see also \citealt{Larson1984,Gnedin1995}). A number of studies have also looked at the angular momentum transport by the global spiral waves in discs \citep{Larson1990,Spruit1987,RRR16,AR18}, in particular those driven by embedded planets \citet{GR01,R02}. In the BL context, \citet{BRS12,BRS13a} explored some global transport characteristics of the acoustic waves, whereas \citet{BRS13b} did the same in the MHD case\footnote{In their global, unstratified MHD simulations of the near-BL region \citet{BQ17} found accumulation of accreted material in a belt-like structure in the BL, spinning at sub-Keplerian velocity. To allow the accreted material to join the star this belt must dissipate somehow, but it is not yet clear how that happens. Nevertheless, even with the belt \citet{BQ17} still found a substantial wave activity in the inner disc, which is what matters for us in this study.}. {Also, \citet{Dittmann2021} studied how the efficiency of transport varies as the spin of the accreting object changes.}

The aim of our present work is to extend the latter studies by looking at the various characteristics  of the wave-driven transport in the vicinity of the BL --- angular momentum and mass fluxes --- in a systematic fashion across the range of $\M$ values. {A distinct feature of this work compared to previous studies} is that we examine the transport properties of individual modes emerging in our simulations and determine their variation across the different types of modes. Another goal is to explore the wave-driven evolution of the inner parts of the accretion disc adjacent to the BL.

This work is organized as follows. After briefly describing in \S\ref{sect:data} the simulation suite on which this study is based, we remind the reader the basics of the mass and angular momentum transport in accretion discs in \S\ref{sect:other_diag}. Our simulation results on the two kinds of transport are described in \S\ref{sect:transport_mass} and \ref{sect:transport_AM}, respectively. We examine the different contributions to the angular momentum budget in \S\ref{sect:transport_AM_terms} and describe the correlation between the angular momentum and mass fluxes found in our simulations in \S\ref{sect:Mdot-CS}. The wave-driven evolution of the inner disc is characterized in \S\ref{sect:disk_evol}. Finally, we discuss and summarize our results in \S\ref{sect:disc} and \ref{sect:sum}, respectively.


\section{Description of the numerical data set}
\label{sect:data}


Our present study is based on the data produced by a set of 2D simulations (39 runs in total) presented in \citetalias{Coleman2021}. These runs used \athena \citep{athena} to solve hydrodynamic equations with the globally isothermal equation of state (EOS), i.e. $P=\Sigma c_s^2$ for a constant sound speed $c_s$, where $P$ is the vertically integrated pressure and $\Sigma$ is the surface density. Magnetic fields and disc self-gravity have been ignored, {and the accretor was initially non-spinning}. 

These simulations are set in polar $r-\phi$ coordinates, with the azimuthal coordinate covering full $2\pi$. \newedit{ In the radial direction the simulation domain starts below the stellar radius $R_\star$ (at $R=(0.608-0.933)R_\star$, depending on $\M$) and extends in the disc out to $4R_\star$, see \citetalias{Coleman2021}. It is important that our simulation domain includes not only the disc and the BL, but also the outer layers of the star to a depth of several hydrostatic scale heights. We apply the "do nothing" boundary conditions (i.e. maintain fluid variables at their initial values) at both radial boundaries.} The radial grid is uniformly spaced in $\log r$ to increase resolution in the vicinity of the BL. Simulations were run for every integer value of $\M$ in the interval $5\le \M\le 15$. For three values of $\M=6,9,12$ we run multiple simulations with the different forms of the initial noise spectrum (triggering sonic instability in the BL) and numerical resolution, to test their impact on the outcomes. 

A distinctive feature of our simulations is their detailed, purpose-built, real-time analysis. They employed a high-cadence sampling of the outputs, allowing a highly informative analysis of the outputs to be performed. In particular, we ran on-the-fly fast-Fourier transforms (FFTs) of various fluid variables, giving us new, previously unobtainable diagnostic capabilities. This allowed us to detect and characterize a number of different modes present in the system, with their distinct azimuthal wavenumbers $m$ and pattern speeds $\Omega_P$, at every moment of time. The ability to automatically detect and analyze different modes present in the system is an important improvement of the study presented in \citetalias{Coleman2021} compared to the existing works. 

In the following, when presenting our results, we will be using units in which $R_\star=1$ and $\Sigma(R_\star)=1$ at $t=0$; in these units the initial profile of the surface density is $\Sigma=r^{-3/2}$. We also set the Keplerian velocity at the surface of the star $v_K(R_\star)$ to be unity, which implies that $c_s=\mathcal{M}^{-1}$ and $GM_\star=1$. This choice makes the Keplerian period at $R_\star$ to be $\tau_{\star}=2\pi$, and we will often express time in the form of $t/2\pi$, or in the units of $\tau_{\star}$.


\subsection{Main findings of Paper I}
\label{sect:findings}


We now recount the main conclusions of the wave morphology study presented in \citetalias{Coleman2021}, that will allow us to better interpret the results of the current work. 

Our suite of simulations reveals a complicated pattern of wave activity in the vicinity of the BL, with a number of different modes operating in the inner disc. In addition to the upper and lower acoustic modes previously described in \citet{BRS13a}, we have also discovered a new type of modes, so called {\it vortex-driven} modes. They appear as global spiral arms extending from the vicinity of the BL into the upper disc, and owe their existence to the localized vortex-like structures that form in the disc next to the BL and launch these density waves. 

Another type of modes that we routinely detect in our runs are the so called {\it resonant} modes (owing their existence to a particular geometric resonance condition, \citealt{BRS12}), which are present only in the disc. Similar to the lower acoustic modes, the resonant modes are trapped between the stellar surface and the inner Lindblad resonance, whose location $r_\mathrm{ILR}$ in a Keplerian disc with $\Omega=\Omega_K=(GM_\star/r^3)^{1/2}$ is given by
\ba   
r_\mathrm{ILR}=R_\star\left[\frac{\Omega_K(R_\star)}{\Omega_P}\frac{m-1}{m}\right]^{2/3}.
\label{eq:r_ILR}
\ea
In \citetalias{Coleman2021} we also found a general tendency of the azimuthal wavenumber $m$ of the most prominent modes to increase with $\M$. These results appear robust with respect to variations of the initial conditions and numerical resolution.


\section{Wave-driven mass and angular momentum transport: basics}
\label{sect:other_diag}


The main goal of the present work is to better understand how matter and angular momentum are transported through the inner disc and the BL, resulting in accretion onto the star. To that effect, we examine how the characteristics of the different modes that we see in our runs are linked to the global evolution of the star-disc system. We do this by exploring the behavior of certain variables derived from our simulations --- mass and angular momentum fluxes --- for different values of $\M$ and connecting them to the changes of the disc properties. 

In all our simulations transport of mass is tracked by measuring the mass accretion rate, 
\begin{align}
    \dot{M}\left(r\right)\equiv-
    r\int_0^{2\pi} v_r\left(r, \phi\right)\Sigma\left(r, \phi\right)\,{\rm d}\phi=-2\pi r\langle \Sigma v_r\rangle,
    \label{eq:Mdot}
\end{align}
{(see Figure \ref{fig:Mdot-split})} where we introduced a shorthand notation 
\ba
\langle f(r)\rangle=(2\pi)^{-1}\int_0^{2\pi}f(r,\phi) d\phi
\ea
for azimuthal averaging of any variable $f(r,\phi)$. Accretion rate is defined such that $\dot M>0$ for {\it inflow} of mass towards the star.

To characterize angular momentum transport we adopt the standard procedure of measuring the total angular momentum flux (AMF) $C_{\rm L}$, defined as
\begin{align}
C_{\rm L} = 2\pi r^2 \ave{\Sigma v_r v_\phi},
\label{eq:C_L}
\end{align}
where $v_r$ and $v_\phi$ are the radial and azimuthal velocity components. By introducing a reference azimuthal velocity of the fluid $v_{\phi,0}(r)$, to be specified later, and the azimuthal velocity perturbation $\delta v_\phi=v_\phi-v_{\phi,0}$, we can further decompose $C_{\rm L}$ into the {\it advective} angular momentum flux $C_{\rm A}$, and the {\it wave} angular momentum flux\footnote{Despite the typographic similarities $c_{\rm s}$ and $C_{\rm S}$ are used to denote two distinct quantities: sound speed and wave-induced angular momentum flux, respectively.} (or Reynolds stress) $C_{\rm S}$, defined as follows: 
\begin{align}
C_{\rm A} &= 2\pi r^2 \ave{\Sigma v_r}v_{\phi,0}=-\dot M ~r v_{\phi,0},
\label{eq:C_A}\\
C_{\rm S} &= 2\pi r^2 \ave{\Sigma v_r \left(v_\phi-v_{\phi,0}\right)}=2\pi r^2 \ave{\Sigma v_r \delta v_\phi},
\label{eq:C_S}
\end{align}
so that $C_{\rm L}=C_{\rm A}+C_{\rm S}$. In our case the stress contribution $C_{\rm S}$ arises primarily because of the oscillatory wave-like motions and is driven by the wave modes rather than some turbulence, as would be the case for e.g. the MRI. 

In this work, when analyzing simulation outputs, we will most often be choosing $v_{\phi,0}$ to be the mean azimuthal velocity of the fluid,
\begin{align}
v_{\phi,0}=\ave{v_\phi},
\label{eq:v_0}
\end{align}
following \citet{Fromang2006} and \citet{Flock2011}, among others. Note that alternative definitions of $C_{\rm S}$ using other choices of $v_{\phi,0}(r)$ are also possible, see e.g. \citet{JU16}, \citet{AR18}, as well as \S\ref{sect:fluxes_rel}. 

Because of oscillatory, intrinsically time-dependent nature of the acoustic mode-driven transport, we find $\dot M$ and the different angular momentum flux contributions to be highly time-dependent. For that reason, we resort to averaging these physical quantities over sufficiently long time intervals to provide meaningful comparison between them. We describe mathematical details of such averaging procedures in Appendix \ref{sect:averaging}.


\subsection{Relation between \texorpdfstring{$\dot M$}{Mdot} and \texorpdfstring{$C_{\rm S}$}{CS}}
\label{sect:fluxes_rel}


The mass and angular momentum fluxes through the disc are closely related to each other, which can be demonstrated quite generally starting from the hydrodynamic equations of motion and continuity. 
In particular, it was shown in \citet{BRS13a} that by choosing $v_{\phi,0}(r)$ in the {\it mass-weighted} form \citep{BAL98,BalPap1999}
\begin{align}
v_{\phi,0}=v_\Sigma\equiv\frac{\ave{\Sigma v_\phi}}{\ave{\Sigma}},
\label{eq:v_0-weighted}
\end{align}
different from that given by the equation (\ref{eq:v_0}), and defining the wave AMF (\ref{eq:C_S}), or Reynolds stress, $C_{\rm S}$ accordingly (i.e. with $v_{\phi,0}=v_\Sigma$), the equation representing conservation of the angular momentum can be cast in the following form:
\begin{align}
\dot M\frac{\partial l}{\partial r} =\frac{\partial C_{\rm S} }{\partial r}+2\pi r^3\langle\Sigma\rangle \frac{\partial \Omega_0}{\partial t}.
\label{eq:AM_terms}
\end{align}
Here $\Omega_0=v_\Sigma/r$ and $l=rv_\Sigma=\Omega_0 r^2$ are the angular frequency and the specific angular momentum corresponding to the reference azimuthal velocity $v_\Sigma$. If instead of (\ref{eq:v_0-weighted}) we adopted $v_{\phi,0}$ in the form (\ref{eq:v_0}), then the last term in the equation (\ref{eq:AM_terms}) would look differently. 

Equation (\ref{eq:AM_terms}) directly relates the transport of mass --- the term proportional to $\dot M$ in the left hand side --- to the transport of the angular momentum in the disc --- the divergence  of the wave angular momentum flux $C_{\rm S}$ in the right-hand side. The second term in the right hand side, usually neglected in studies of accretion processes, arises in discs which evolve sufficiently rapidly for their mean angular frequency $\Omega_0$ to change in time. If we neglect this contribution to the angular momentum balance for a moment, then we find
\begin{align}
\dot M =\left(\frac{\partial l}{\partial r}\right)^{-1} \frac{\partial C_{\rm S} }{\partial r}=\frac{\partial C_{\rm S} }{\partial l}.
\label{eq:AM_terms1}
\end{align}
Note that this expression is fully general and applies for any choice of $v_{\phi,0}$; a particular form of $v_{\phi,0}$ affects the definition of both $C_{\rm S}$ and $l$.

In truly viscous discs with kinematic viscosity $\nu$ one should use viscous angular momentum flux $F_J=-2\pi\nu\Sigma r^3 d\Omega/dr$ instead of $C_{\rm S}$ \citep{Lynden1974}. In particular, in discs with radially constant $\dot M$ one finds $F_J=\dot M l$.

However, as we will demonstrate in \S\ref{sect:transport_AM_terms}, the second term in the right hand side of equation (\ref{eq:AM_terms}) often plays an important role near the BL. This contribution to the angular momentum balance has been previously considered in \citet{BRS13a}, and its relation to the evolution of the disc properties was explored in \citet{AR18}.


\section{Transport of mass}
\label{sect:transport_mass}


We start by describing our results for the mass flux $\dot M$. According to equation (\ref{eq:C_A}), $\dot M$ is closely related to $C_{\rm A}$. Indeed, in the disc $v_{\phi,0}$ is typically pretty close to the Keplerian velocity (see \S\ref{sect:omega_evol}), allowing one to approximate $C_{\rm A}\approx -\dot M l_K$ outside the BL, where $l_K(r)=\sqrt{GM_\star r}$ is the specific angular momentum for a Keplerian disc. The profile of $C_{\rm A}(r)$ is plotted in Figures \ref{fig:AM1}-\ref{fig:AM2} (blue curve in the middle row of each individual subpanel) for simulations at varied $\M$ and at different moments of time. It is discussed in more details in \S\ref{sect:transport_AM_types}, but for now we will mention some key features pertinent for the $\dot M$ behavior. 

First, in all plotted cases $C_{\rm A}(r)$ (and $\dot M$) exhibits a deep minimum (maximum) close to the BL. This is naturally explained by the enhanced wave activity near the BL, and results in important consequences for the surface density evolution in the inner disc discussed in \S\ref{sect:disk_evol}. \newedit{ That $\dot M$ is not constant in radius is due to the fact that in our runs mass accretion is caused only by the wave activity near the star and inflow of mass from large radii is absent.} Second, in most cases both $C_{\rm A}(r)$ and $\dot M$ substantially diminish in amplitude far from the accretor. We will relate this behavior to the angular momentum fluxes carried by the different types of waves in \S\ref{sect:transport_AM}.


\subsection{\texorpdfstring{$\dot M$}{Mdot} decomposition}
\label{sect:Mdot_decompose}

\begin{figure}
	\includegraphics[width=\linewidth]{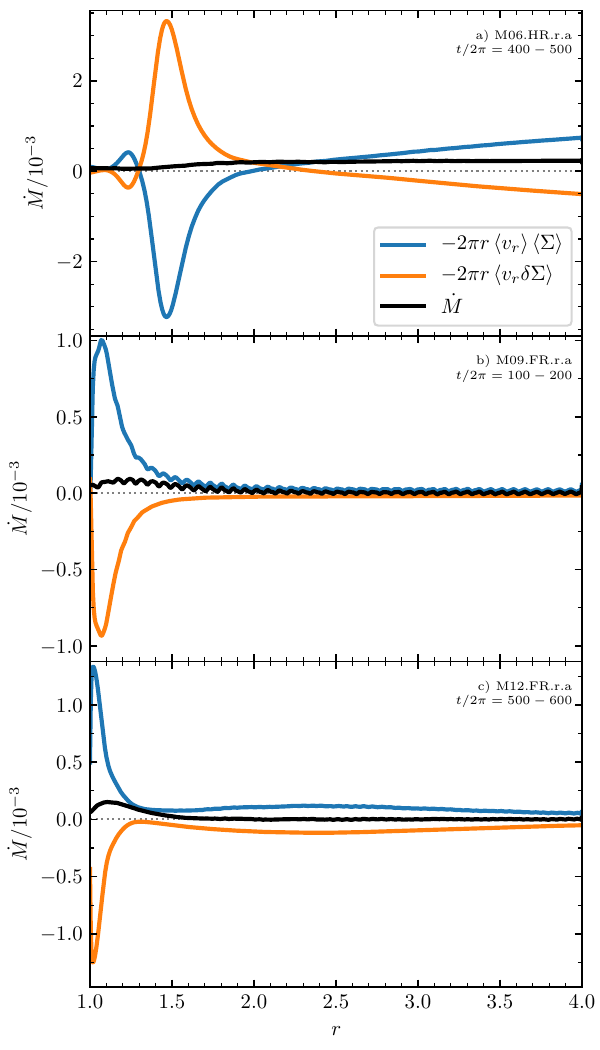}
    \caption{Different contributions to the total mass accretion rate $\dot M$ (black curve) in equation (\ref{eq:Mdot1}), plotted as a function of $r$ for several simulations (see \citetalias{Coleman2021} for their naming notation). The advective term $2\pi\langle \Sigma v_r\rangle\langle v_r\rangle$ is shown in blue, while the diffusive term $-2\pi\langle \delta\Sigma v_r\rangle$ is the orange curve. Averaging over the interval $t/2\pi=100-200$ is performed.}
    \label{fig:Mdot-split}
\end{figure}

Our simulations reveal an interesting feature of the $\dot M$ behavior near the BL, that we discuss next. Let us write down $\Sigma(r,\phi)=\langle\Sigma\rangle +\delta\Sigma(r,\phi)$, where $\delta\Sigma(r,\phi)$ is the perturbation of the surface density relative to its azimuthally-averaged value $\langle\Sigma\rangle$. Then the definition (\ref{eq:Mdot}) can be cast as
\begin{align}
    \dot{M}\left(r\right)=-2\pi r\langle \Sigma \rangle\langle v_r\rangle-2\pi r\langle \delta\Sigma v_r\rangle,
    \label{eq:Mdot1}
\end{align}
illustrating the separation of $\dot M$ into two distinct contributions --- the one due to the advection of the mean surface density $\langle\Sigma\rangle$ (first term) and the contribution due to the correlation between the non-axisymmetric fluctuation of $\Sigma$ and $v_r$. We illustrate the behavior of these contributions in Figure \ref{fig:Mdot-split} for three simulations with different values of $\M$. One can see that in all cases both contributions to $\dot M$ are, in general, equally important: they are comparable in magnitude and tend to offset each other so that the total $\dot M$ is considerably lower in magnitude than each of these terms over large radial ranges (which is especially noticeable near the BL). For that reason, one cannot directly connect $\dot M(r)$ to the radial profiles of $\langle v_r\rangle$ and $\langle\Sigma\rangle$. 

Importance of the $\dot M$ contribution due to the correlation between $\delta\Sigma$ and $v_r$ is one of the distinctive features of the global, acoustic wave-driven transport. In the classical picture of laminar viscous disc accretion \citep{SS73,Lynden1974} this contribution is identically zero since the disc is axisymmetric and $\delta\Sigma=0$. In our case, this contribution is non-zero in the vicinity of the BL because of highly correlated $\delta\Sigma$ and $v_r$ for wave-like fluid motions and the nonlinear dissipation of these waves, which leads to damping of the angular momentum flux carried by the waves. Interestingly, $\langle \delta\Sigma v_r\rangle$ is often more regular than $\langle \Sigma \rangle\langle v_r\rangle$, since the latter sometimes exhibits small-scale spatial variability even upon the long-term time averaging, see Figure \ref{fig:Mdot-split}b. 

We are not aware of any studies mentioning the role of the mass transport term proportional to $\langle \delta\Sigma v_r\rangle$ in simulations featuring turbulence, self-consistently driven by a local hydrodynamic (e.g. the vertical shear instability) or MHD (e.g. MRI) mechanism. Examining this issue may be interesting for figuring out whether in our case the non-zero second term in equation (\ref{eq:Mdot1}) arises due to the intrinsically global nature of the wave-driven transport near the BL (see \S\ref{sect:global}), or it is also present when accretion is mediated by local processes.

\section{Angular momentum transport}
\label{sect:transport_AM}

\begin{figure*}
    \vspace*{-1em}
	\includegraphics[width=\linewidth]{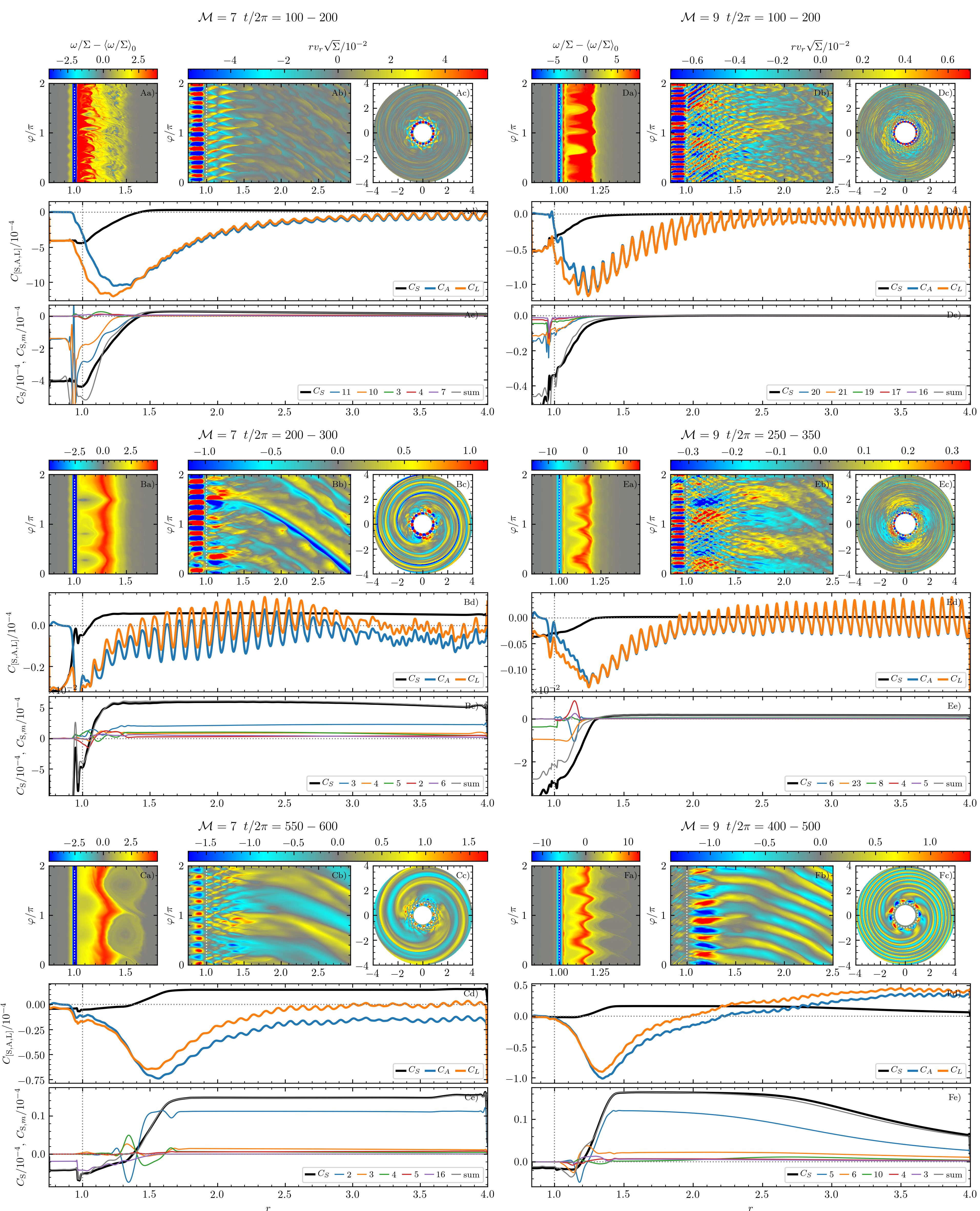}
	\vspace*{-2em}
    \caption{Figure illustrating angular momentum flux data for simulations with $\M=7$ (run M07.FR.r.a, left panels) and $\M=9$ (run M09.FR.r.a, right panels). Each panel (labeled with capital letters) consists of 5 sub-panels, labeled by their subscript and described in the beginning of \S\ref{sect:transport_AM}. Every panel is associated with a time interval over which the AMF data are computed ($t/2\pi$, shown above the panel) and a representative moment of time $T/2\pi$ within this interval, indicated as follows. 
    }
    \label{fig:AM1}
\end{figure*}

We now turn to the details of the angular momentum transport in the vicinity of the BL. Figures \ref{fig:AM1},\ref{fig:AM2} illustrate the behavior of various representative variables during three different periods of time for the four representative simulations with $\M=7,9,12$ and $15$. The morphological description of the modes emerging in two of the runs --- for $\M=9$ (M09.FR.r.a) and $\M=15$ (M15.FR.r.a) --- has been provided in \citetalias{Coleman2021}, whereas the $\M=7$ (M07.FR.r.a) and $\M=12$ (M12.FR.mix.a) runs have not been discussed in detail in \citetalias{Coleman2021}.  

For each $\M$ we select three time intervals of length $50$ or $100$ orbits and perform averages (as described in Appendix \ref{sect:averaging}) of various angular momentum fluxes --- total $C_{\rm L}$, advective $C_{\rm A}$, and wave $C_{\rm S}$, defined by the equations (\ref{eq:C_L})-(\ref{eq:C_S}) with $v_{\phi,0}=\ave{v_\phi}$ given by the equation (\ref{eq:v_0}). Their radial profiles are shown in the middle row of each panel. As mentioned earlier in \S\ref{sect:transport_mass}, the behavior of the advective flux $C_{\rm A}$ in the disc is closely related to the radial profile of the mass accretion rate $\dot M$. 

We also select a representative moment of time within each interval, which illustrates a set of modes typical for that period, and show snapshots of various fluid variables in the top row the panel corresponding to that time interval. The format is the same as in Figure 2 of \citetalias{Coleman2021}, namely, we show a polar map of the wave amplitude variable $r v_r\sqrt{\Sigma}$ (right), a Cartesian map of the same variable over the reduced radial range to highlight the near-BL details (middle), and a Cartesian map of the vperturbation of ortensity $\zeta=\omega/\Sigma$ (here $\omega=\nabla\times \mathbf{v}$ is the vorticity) relative to its initial value near the BL (left).

Finally, the bottom row shows the decomposition of the wave angular momentum flux $C_{\rm S}$ into the contributions provided by the individual Fourier modes of the fluid perturbation, which are dominant through the plotted time interval (in a time integrated sense). To second order in perturbed variables the Fourier contribution $C_{{\rm S},m}$ to $C_{\rm S}$ from the $m$-th mode is \citep{BRS13a}
\begin{align}
    C_{{\rm S},m}=2\pi r^2 \ave{\Sigma}\left(v_{r,m} v_{\phi,m}^\ast+v_{r,m}^\ast v_{\phi,m}\right),
\label{eq:C_S_m}
\end{align}
where the asterisks denote complex conjugates, so that $C_{\rm S}\approx \sum_{m=1}^\infty C_{{\rm S},m}$. The particular $C_{{\rm S},m}$ profiles shown in the figure are for the dominant modes chosen according to the procedure described in Appendix \ref{sect:averaging}. The gray curves, representing the sum of only the dominant harmonics indicated in each panel, sometimes show deviations from the full $C_{\rm S}$ (black), computed using equation (\ref{eq:C_S}). These differences arise because a number of other, less significant modes not shown in these plots also contribute to the full $C_{\rm S}$, and also because equation (\ref{eq:C_S_m}) is only second order accurate in fluid perturbations.

The decomposition of $C_{\rm S}$ into contributions from the different azimuthal harmonics allows us to explore the role played by the individual modes (which, once their $m$ is known, can be easily associated with the different types of waves) in transporting the angular momentum in the vicinity of the BL. {This is the key advantage of our present work compared to \citet{BRS13a} and \citet{Dittmann2021}, who did not perform such decomposition. Another improvement is the larger range of Mach numbers explored in our study, providing us with a better understanding of the dependence of transport properties on $\M$.}


\subsection{Different AMF contributions}
\label{sect:transport_AM_types}

\begin{figure*}
    \vspace*{-1em}
	\includegraphics[width=\linewidth]{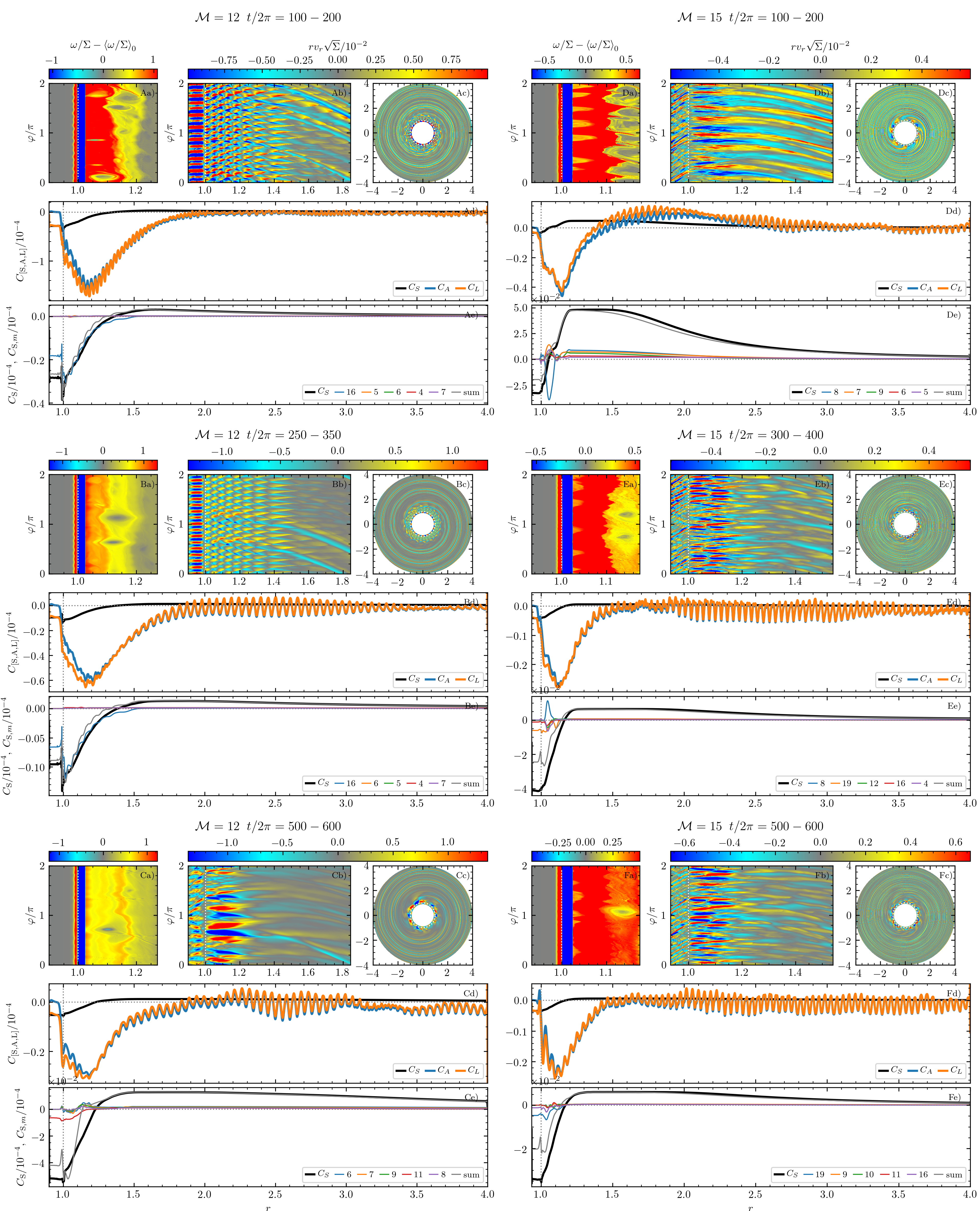}
	\vspace*{-2em}
    \caption{Same as Fig. \ref{fig:AM1} but for $\M=12$ (run M12.FR.mix.lc.a, left panels) and $\M=15$ (run M15.FR.r.a, right panels). For $\M=12$: (A) $T/2\pi=150$, $t/2\pi=100-200$; (B) $T/2\pi=300$, $t/2\pi=250-350$; (C) $T/2\pi=575$, $t/2\pi=550-600$. For $\M=15$: (D)  $T/2\pi=150$, $t/2\pi=100-200$; (E) $T/2\pi=350$, $t/2\pi=300-400$; (F) $T/2\pi=550$, $t/2\pi=500-600$.
    }
    \label{fig:AM2}
\end{figure*}

Middle sub-panels of Figures \ref{fig:AM1}-\ref{fig:AM2} display the advective angular momentum flux $C_{\rm A}$, the wave flux $C_{\rm S}$, and their sum $C_{\rm L}$ on the same scale. A notable feature of the $C_{\rm A}$ curves is the small-scale oscillations that they often exhibit. These oscillations are likely caused by the intrinsic time-variability, since the inner disc is pervaded by the multiple wave modes. This variability ends up manifesting itself in the spatial domain, even though we perform time averaging over a substantial interval of time, see Appendix \ref{sect:averaging}.

One can see that $|C_{\rm A}|$ peaks not too far from the stellar surface and its maximum amplitude is always larger than that of $C_{\rm S}$ by a factor of several. This makes sense since, to zeroth order (certainly in steady state), $C_{\rm A}\propto\dot M \propto \partial C_{\rm S}/\partial r$, see equations (\ref{eq:C_A})), (\ref{eq:AM_terms}). Since $C_{\rm S}$ typically varies over a narrow region near the BL, its radial derivative can reach high values. Naturally, the narrower is the near-BL region over which $C_{\rm S}$ varies, the larger should we expect the difference in amplitudes between $C_{\rm A}$ and $C_{\rm S}$ to be. And indeed, in Figure \ref{fig:AM1} (for $\M=7$) we find $C_{\rm S}$ variation in the range $1< r\lesssim 1.5$ to result in the ratio of the maximum amplitudes $C_{\rm A}$ and $C_{\rm S}$ being $\sim (2-2.5)$, while in Figure \ref{fig:AM2}Fd (for $\M=15$) $C_{\rm S}$ variation in the range $1< r\lesssim 1.2$ leads to the maximum ratio of fluxes $\sim 5-6$.

In most cases we find $C_{\rm A}$ to be negative, as expected for mass inflow with $\dot M>0$. Notable exceptions include $\M=9$ run at $t/2\pi=400-500$ (Figure \ref{fig:AM1}Fd) and $\M=15$ run at $t/2\pi=100-200$ (Figure \ref{fig:AM2}Dd), for which $C_{\rm A}>0$ (i.e. $\dot M<0$) at large distances. This can again be understood on the basis of the equations (\ref{eq:AM_terms}) or (\ref{eq:AM_terms1}), by noticing that in these cases $C_{\rm S}$ is {\it positive and decays} with $r$ far from the star as a result of dissipation {(it will become clear in \S\ref{sect:transport_AM_modes},\ref{sect:dif-modes} that this behavior is caused by the nonlinear damping of the vortex-driven modes)}. As a consequence, $\partial C_{\rm S}/\partial r<0$ there, leading to $\dot M<0$ (outflow) and positive $C_{\rm A}$.

\subsection{Wave AMF \texorpdfstring{$\CS$}{CS} as a function of \texorpdfstring{$\M$}{M} and \texorpdfstring{$m$}{m}}
\label{sect:transport_AM_modes}


We now describe the behavior of the wave AMF $C_{\rm S}$ for different values of $\M$ shown in Figures \ref{fig:AM1},\ref{fig:AM2} in conjunction with the morphological characteristics of the different waves present in the system.


\subsubsection{$\M=7$ run}
\label{sect:M=7}

We start our description of the $\CS$ behavior with the $\M=7$ run,
shown in Figures \ref{fig:AM1}A-C. At early times\footnote{{Note that by that time a significant depression in surface density has already formed in the inner disc, see Figure \ref{fig:multi_st}a.}} ($t/2\pi=125$) the inner disc is dominated by the $m=11$ lower mode trapped within $r\lesssim 1.5$ (panel Ab), which roughly corresponds to the $r_{\rm ILR}$ of this mode (its $\Omega_P\approx 0.46$). This mode gives rise to the negative AMF $C_{\rm S,11}$, see the blue curve\footnote{The orange curve in that panel corresponding to the $m=10$ lower mode also provides a substantial contribution to $\CS$ near the star. This mode is not seen in panels (Aa)-(Ac) because the data shown in panels (Ad), (Ae) are integrated over an extended time interval and $m=10$ signal gets washed out.} in panel (Ae). As a result, $\CS$ starts negative at $r=R_\star$ and crosses zero around $r_{\rm ILR}$. 

Outside the resonant cavity there is a set of rather incoherent vortex-driven spiral arms, excited by a number of chaotic vortices residing at $r\lesssim 1.5$, visible in panel (Aa). These weak arms drive low-amplitude but {\it positive} $C_{\rm S}$ outside $r=1.5$, see panel (Ad).

By $t/2\pi=225$ the amplitude of the wave activity decreases; $m=11$ lower mode gets replaced with the lower $m=10$ mode, which is prominent inside the star but is rather weak in the disc, see panel (Bb). As a result, the magnitude of the (negative) segment of the $C_{\rm S}$ curve in the inner disc drops by about two orders of magnitude, and it gets confined close to the star (panel (Bd)). Also by that time, vortensity evolution produces three prominent vortices at $r\approx 1.1$ (panel (Ba)), which excite strong vortex-driven spirals in the outer disc, increasing the amplitude of $C_{\rm S}>0$ in the outer disc (panel (Bd)).

In the final stages of this $\M=7$ simulation, at $t/2\pi=575$, a lower $m=16$ mode clearly shows in the disc at $r\lesssim 1.8$ (although inside the star $m=8$ mode dominates), although with rather modest amplitude (panel (Cb)). This, again, makes $\CS$ negative close to the star, see the red curve for $C_{{\rm S},16}$ in panel (Ce). Also, the vortensity map in panel (Ca) reveals two strong, azimuthally elongated vortices centered at $r\approx 1.5$. These "rolls", as we termed them in \citetalias{Coleman2021}, drive a pair of azimuthally extended vortex-driven spiral arms with $m=2$ in the outer disc. Their $C_{{\rm S},2}>0$ makes $\CS$ positive in the outer disc (see panel (Ce)), with the amplitude $\approx 2.5$ times higher than at $t/2\pi=225$.

Already this run alone demonstrates that depending on a type of the mode dominating in a particular region of the disc, one can get rather different behaviors of $\CS$. Our subsequent description of the runs for other values of $\M$ will reinforce this conclusion.


\subsubsection{$\M=9$ run}
\label{sect:M=9}

The $\M=9$ run in its early phases $t/2\pi=150$ reveals a set of lower $m=19-21$ modes confined within $r_{\rm ILR}\approx 2.2$, corresponding to their $\Omega_P\approx 0.3$, see Figure \ref{fig:AM1}Db,Dc. These modes give rise to $\CS<0$ near the star. Similar picture persists also at $t/2\pi=275$, see panels (Eb)-(Ee). At the same time, in the outer disc $\CS$ stays close to zero: unlike the late stages of the $\M=7$ run, coherent vortices do not emerge near the star (see panels (Da), (Ea)), so vortex-driven spirals are very weak.

An interesting feature of the $\M=9$ run pointed out in \citetalias{Coleman2021}, is the emergence of $m=2$, radially elongated mode with very low pattern speed $\Omega_P\approx 0.15$. It is readily noticeable in panel (Eb), however its signal is not present in panel (Ee), which means that its AMF contribution $C_{{\rm S},2}$ is negligible. This is hardly surprising since this mode has radial wave number $k_r=0$.

Finally, at $t/2\pi=150$ the picture changes dramatically, see panels (Fb),(Fc). Near the star, for $r\lesssim 1.4$, the disc is dominated by the $m=6$ resonant mode (see \citetalias{Coleman2021} for details), whereas in the outer disc there are global $m=5$ spiral arms driven by a set of five vortex rolls located at $r\approx 1.3$, see panel (Fa). The resonant mode gives rise to a small (but negative near the star) AMF contribution $C_{{\rm S},6}$, whereas the vortex-driven modes produce $C_{{\rm S},5}>0$ in the outer disc, similar to $\M=7$ run.


\subsubsection{$\M=12$ run}
\label{sect:M=12}

An $\M=12$ run that we use in this work is different from the run employed in \citetalias{Coleman2021}. This is done in part to illustrate the fact, pointed out in \citetalias{Coleman2021}, that all $\M=12$ runs look very similar to each other, in a much more homogeneous way than for any other value of $\M$. 

Indeed, comparing Figure \ref{fig:AM2}A-C with the Fig. 7 of \citetalias{Coleman2021}, one can see a very familiar pattern: for most of the simulation, including $t/2\pi=150$ and $300$ shown in Figure \ref{fig:AM2}A,B, the inner disc (out to $r\approx 1.6$) is dominated by the lower $m=16$ mode, accompanied by a set of narrow vortex-driven spirals clearly visible in the outer disc. As expected, this pattern results in $C_{{\rm S},16}$ driving $\CS<0$ near the star, with $\CS$ becoming positive further out, at $r\gtrsim 1.3-1.5$, see panels (Ae), (Be).

Closer to the end of the run, at $t/2\pi=550$, one can see a combination of a resonant and a lower $m=11$ modes driving $\CS$ below zero near the star, with the weak vortex-driven modes carrying $\CS>0$ outside $r\approx 1.3$.


\subsubsection{$\M=15$ run}
\label{sect:M=15}

The $\M=15$ simulation is the highest $\M$ run carried out as a part of our simulation suite described in \citetalias{Coleman2021}. Unlike other simulations described so far, this run shows quite substantial upper acoustic mode activity inside the star \citep{BRS13a}, see Figure \ref{fig:AM2}Db,Eb,Fb, although the manifestations of this mode type in the disc are not as clear. 

In this run we witness the appearance of 7-8 vortex rolls at $r\approx 1.15$ rather early on, already by $t/2\pi=150$, see panel Da. These rolls drive a set of global, relatively narrow spiral arms propagating in the outer disc and making $\CS$ positive beyond $r\approx 1.1$. For $1<r\lesssim 1.1$ we find $\CS<0$, which could be due to the resonant modes starting to develop in this part of the disc.

These resonant modes do become rather prominent (and more radially extended) at later time, e.g. at $t/2\pi=350$ and $550$, see panels (Eb),(Fb). They keep $\CS$ negative within $r\approx 1.2$, see panels (Ee),(Fe). At the same time, the vortex rolls appearing earlier in the simulation merge into two (at $t/2\pi=350$) and then one vortex (at $t/2\pi=550$), see panels (Ea), (Fa). They give rise to the vortex-driven mode activity in the outer disc, which maintains positive $\CS$ there, albeit at a reduced amplitude compared to $t/2\pi=150$.


\subsection{Transport by the different types of modes}
\label{sect:dif-modes}


Examination of the bottom sub-panels in all panels of Figures \ref{fig:AM1}-\ref{fig:AM2} reveals that at any moment of time there is often only one or two modes in the disc that dominate the wave AMF $C_{\rm S}$. For example, in the $\M=7$ run at $t/2\pi=100-200$ we find two lower modes with $m=10,11$ to provide the dominant contribution to $C_{\rm S}$ for $r<1.4$ (Figure \ref{fig:AM1}Ae), while at $t/2\pi=550-600$ the vortex-driven $m=2$ mode dominates $C_{\rm S}$ for $r>1.4$ (Figure \ref{fig:AM1}Ce); in the $\M=12$ run the lower $m=16$ mode dominates $C_{\rm S}$ for $r<1.5$ at both $t/2\pi=100-200$ and $250-350$ (Figure \ref{fig:AM2}Ae,Be). As these dominant modes have a particular sign of $C_{\rm S}$, they affect disc evolution in a certain way, as described later in \S\ref{sect:disk_evol}. 

A wave propagating in the disc carries angular momentum $\Delta J_{\rm w}\propto [\Omega_P-\Omega(r)]$ (see below), so that the {\it flux} of angular momentum associated with it is
\ba
C_{\rm S}\propto k_r\Delta J_{\rm w}\propto k_r[\Omega_P-\Omega(r)], 
\label{eq:CSwave}
\ea
where $k_r$ is the radial wavenumber. In particular, an outgoing ($k_r>0$) wave has $C_{\rm S}>0$ in the region where $\Omega_P>\Omega(r)$ (e.g. a vortex-driven mode far from the star), whereas an incoming ($k_r<0$) wave has $C_{\rm S}>0$ in the region where $\Omega_P<\Omega(r)$ (e.g. a lower mode reflected off the ILR). This is what one also finds for the density waves excited by planets, outside and inside of the planetary semi-major axis. On the other hand, in the disc region where $\Omega_P<\Omega(r)$ an outgoing ($k_r>0$) wave has $C_{\rm S}<0$ (e.g. a lower mode propagating away from the star). 

Nonlinear wave damping provides an intuitive way to understand why $\Delta J_{\rm w}\propto {\rm sgn}[\Omega_P-\Omega(r)]$. Let us consider a shock wave with a pattern speed $\Omega_P$. Whenever $\Omega_P>\Omega(r)$ (which is true far from the accretor) the shock overtakes the disc fluid, {\it accelerating} it in the azimuthal direction upon crossing the shock as a result of the shock jump conditions. As a result, disc fluid receives {\it positive} $\Delta J$ from the wave. On the contrary, for $\Omega_P<\Omega(r)$ (fulfilled close to the BL, in the inner disc) the disc fluid catches up with the shock front and {\it decelerates} upon crossing it; this means that the wave has transferred {\it negative} angular momentum to the disc, i.e. that $\Delta J_{\rm w}<0$.


\subsubsection{Transport by the lower acoustic modes}
\label{sect:transport-lower}


Lower modes are trapped inside the resonant cavity at $1<r<r_{\rm ILR}$, and in this part of the disc $\Omega_P<\Omega(r)$. The mode is excited in the BL \citep{BRS13a} and first propagates out (i.e. $k_r>0$) towards the ILR. At the ILR (where its $k_r\to 0$) it gets reflected and propagates back to the stellar surface with $k_r<0$ (see Figure \ref{fig:AM2}Ab for an illustration). {As a result,} the outgoing lower mode has {angular momentum flux} $C_{\rm S}^{\rm out}<0$, whereas the incoming (reflected) one carries $C_{\rm S}^{\rm in}>0$. The full angular momentum flux is the sum of the two,
\ba  
C_{\rm S}(r)=C_{\rm S}^{\rm out}(r)+C_{\rm S}^{\rm in}(r).
\label{eq:full-lower-CS}
\ea  

Because of the nonlinear wave damping, $|C_{\rm S}^{\rm out}(r)|$ gradually decreases in amplitude as $r$ increases, whereas $|C_{\rm S}^{\rm in}(r)|$ decreases in amplitude with decreasing $r$. As a result, for any $1<r<r_{\rm ILR}$ one finds that the negative $C_{\rm S}^{\rm out}(r)<C_{\rm S}^{\rm out}(r_{\rm ILR})$, whereas the positive $C_{\rm S}^{\rm in}(r)<C_{\rm S}^{\rm in}(r_{\rm ILR})$. Also, $C_{\rm S}^{\rm out}(r_{\rm ILR})=-C_{\rm S}^{\rm in}(r_{\rm ILR})$ since $C_{\rm S}(r_{\rm ILR})=0$. With this in mind, equation (\ref{eq:full-lower-CS}) can be cast as
\ba  
C_{\rm S}(r)=\left[C_{\rm S}^{\rm out}(r)-C_{\rm S}^{\rm out}(r_{\rm ILR}\right]+\left[C_{\rm S}^{\rm in}(r)-C_{\rm S}^{\rm in}(r_{\rm ILR})\right]<0,
\label{eq:full-lower-CS1}
\ea  
showing that $\CS$ of the lower modes is always negative, in agreement with Figures \ref{fig:AM1}-\ref{fig:AM2}. As a result, these modes drive mass flow towards the accretor, $\dot M>0$, see  \S\ref{sect:disk_evol}.


\subsubsection{Transport by the resonant modes}
\label{sect:transport-resonant}


Resonant modes are very similar to the lower modes, since they are also trapped in the resonant cavity, where $\Omega_P<\Omega(r)$. They also propagate both towards and away from the ILR. As a result, by applying the same logic as in \S\ref{sect:transport-lower}, we conclude that resonant modes have $C_{\rm S}(r)<0$ and their dissipation gives rise to $\dot M>0$.

Typically, the magnitude of $C_{\rm S}$ for the resonant modes is lower than $|C_{\rm S}|$ of the lower modes. This is caused by the smaller azimuthal wavenumber $m$ of the resonant modes: lower $m$ implies azimuthally wider wave profile, slower nonlinear wave steepening and shocking, and weaker dissipation. As a result, for resonant modes $C_{\rm S}^{\rm out}(r)$ and $C_{\rm S}^{\rm in}(r)$ in equation (\ref{eq:full-lower-CS}) are closer in amplitude to each other (but still different in sign) than for the lower modes, resulting in their more substantial cancellation and lower $|C_{\rm S}|$.


\subsubsection{Transport by the vortex-driven modes}
\label{sect:transport-vortex}


Vortex-driven modes have $\Omega_P$ set by the angular frequency of their parent vortices, which is typically high since the vortices reside not too far from the BL. As a result, these modes propagate with $k_r>0$ and $\Omega_P>\Omega(r)$ in the outer disc, {\it outside} the corotation region of the mode. This means that these modes have $C_{\rm S}>0$ and (in steady state) result in $\dot M<0$, i.e. mass {\it outflow} in the outer disc. 

As shown in \citetalias{Coleman2021}, vortex-driven waves typically fall into two classes: those produced by the compact, isolated vortices (e.g. see Figure \ref{fig:AM1}Ba,Bb,\ref{fig:AM2}Aa,Ab), and the ones driven by the more azimuthally elongated "rolls" (e.g. see Figure \ref{fig:AM1}Ca,Cb,Fa,Fb). The former have small azimuthal width and are superpositions of a number of high-$m$ perturbation modes (see \citetalias{Coleman2021}). As a result, their constituent modes are often not singled out individually in our plots, see e.g. the $C_{\rm S}>0$ segments in the outer disc in Figures  \ref{fig:AM1}Be,\ref{fig:AM2}Ae, which are not dominated by a mode with a single value of $m$. On the contrary, the azimuthally extended rolls typically produce waves with a well defined, low value of $m$. As a result, in such cases a particular $C_{{\rm S},m}$ dominates the full $C_{\rm S}$ in the outer disc, see Figures \ref{fig:AM1}Ce,Fe. 

The way in which vortex-driven modes dissipate depends on their amplitude, as well as $\M$ and $m$. Higher $m$ accelerates wave evolution into a shock because of the smaller azimuthal scale. For that reason, the vortex-driven $m=2$ mode in the $\M=7$ run shown Figure \ref{fig:AM1}Ce experiences little damping (its $C_{{\rm S},2}$ is essentially constant in the outer disc), whereas the $m=5$ vortex-driven mode in the $\M=9$ run shown Figure \ref{fig:AM1}Fe starts damping appreciably outside $r\approx 2.5$. Note that these modes have similar $\M$ and amplitude, so it is the difference in their $m$ that is responsible for the difference in their decay. Also, everything else being equal, higher amplitude and higher $\M$ \citep{GR01,R02} of a wave promote its faster shocking and dissipation.


\subsubsection{Transport by the upper acoustic modes}
\label{sect:transport-upper}


Upper acoustic modes show up in the disc early on in low $\M$ runs. At high $\M$ they also operate inside the star over long periods of time, see \citetalias{Coleman2021}. These modes are evanescent near the BL, where their $\Omega_P<\Omega(r)$, but become propagating outside their Outer Lindblad Resonance, where their $\Omega_P>\Omega(r)$. In that regard, they are similar to the vortex-driven modes considered above, meaning that they promote mass outflow in the disc, $\dot M<0$. However, the amplitude of the $\dot M$ driven by these modes in the disc is typically too low to make them a significant agent in the disc evolution.


\subsubsection{Transport by other modes}
\label{sect:transport-other}


Another interesting mode type that we see in our simulations is the low $m=2$ mode in $\M=9$ runs clearly visible in Figure \ref{fig:AM1}Eb. This mode has a characteristic radially elongated perturbation pattern (with a clear shift of phase around $r\approx 1.5$) with $k_r=0$. This implies that this mode propagates purely azimuthally and thus does not transport angular momentum in the radial direction. This conclusion is supported by the lack of $C_{{\rm S},2}$ among the different $C_{{\rm S},m}$ appearing in Figure \ref{fig:AM1}Ee.


\subsection{Effective transport coefficients}
\label{sect:alphas}


\begin{figure*}
	\includegraphics[width=\linewidth]{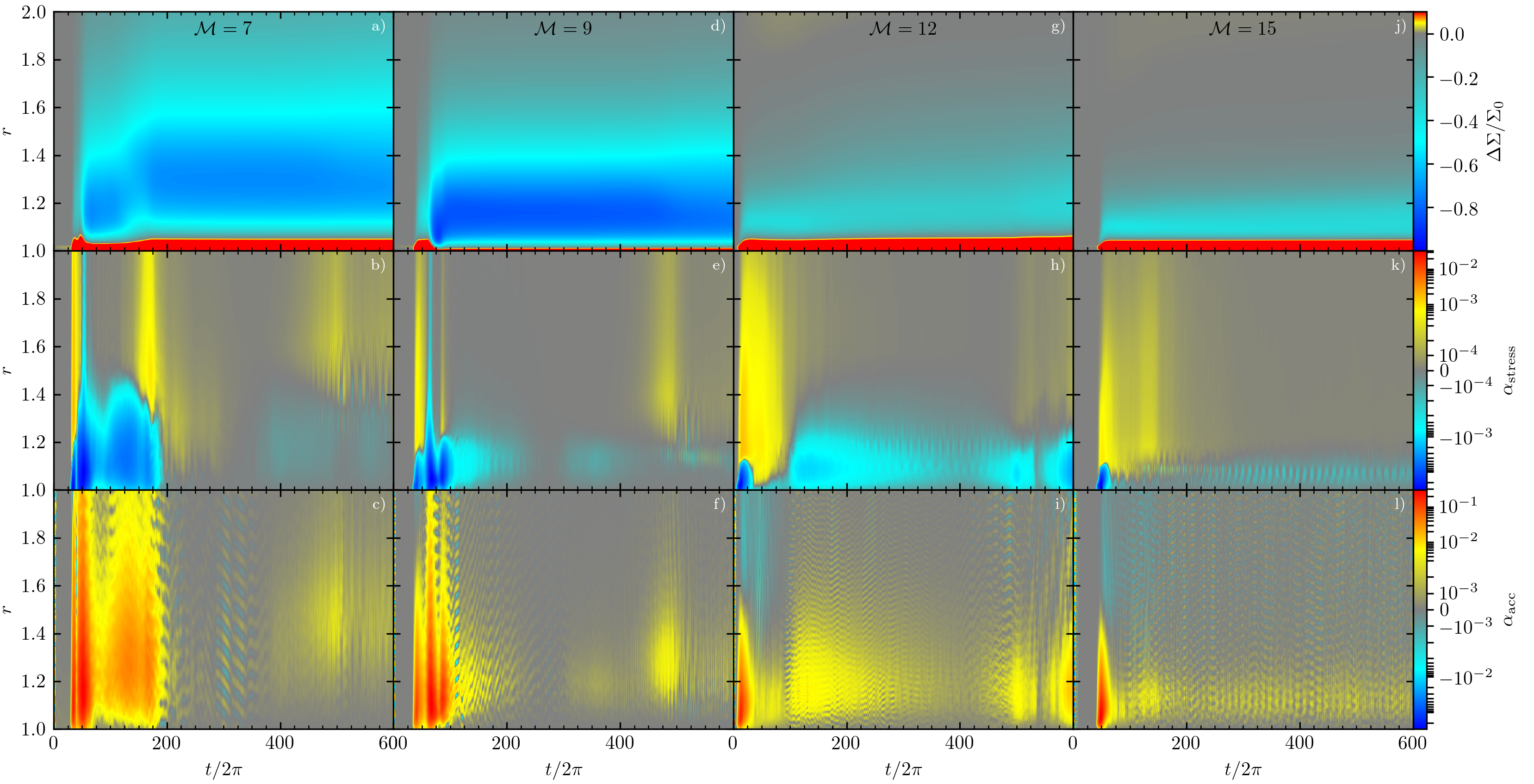}
    \caption{Space-time diagrams of $(\Sigma(t)-\Sigma(0))/\Sigma(0)$ (top), $\alpha_{\rm stress}$ (middle), and $\alpha_{\rm acc}$ (bottom), defined by equations (\ref{eq:alpha_stress})-(\ref{eq:alpha_acc}) for the four simulations described in \S\ref{sect:transport_AM}. Note that the two $\alpha$ parameters often have different signs, even though their amplitudes vary in similar fashion (with 'bursts' over the same time intervals). Large values of $\alpha$ typically lead to significant depletion of material from the disc. See text for details.}
    \label{fig:multi_st}
\end{figure*}

A standard way of quantifying transport processes in an accretion disc is through the use of the dimensionless $\alpha$ parameter, which effectively normalizes stress by the thermal pressure $\Sigma c_s^2$ \citep{SS73}. If the stress is caused by some form of local shear viscosity (e.g. due to MRI turbulence), then the same $\alpha$ also characterizes mass accretion through the disc. However, the transport of mass and energy associated with the acoustic modes is intrinsically non-local as sonic waves are dissipated far from their launching site. 
Nevertheless, we can still formally define the two dimensionless quantities based on Reynolds stress associated with the wave angular momentum flux $C_{\rm S}$ and accretion mass flux $\dot M$ as follows:
\begin{align}
    \alpha_{\rm stress}&\equiv \dfrac{C_{\rm S}}{2\pi r^2 \Sigma c_s^2} = \dfrac{C_{\rm S}}{2\pi r^2 \Sigma}
    \left[\dfrac{\mathcal{M}}{v_K(R_\star)}\right]^2,
    \label{eq:alpha_stress}\\
    \alpha_{\rm acc}&\equiv \dfrac{\dot{M}}{2\pi \Sigma c_s^2} \Omega = \dfrac{\dot{M}}{2\pi \Sigma} \left[\dfrac{\mathcal{M}}{v_K(R_\star)}\right]^2\Omega.
    \label{eq:alpha_acc}
\end{align}
Using equation (\ref{eq:AM_terms1}) we can also write $\alpha_{\rm acc}$ as
\begin{align}
\alpha_{\rm acc} &= \dfrac{\Omega}{2\pi \Sigma} \left[\dfrac{\mathcal{M}}{v_K(R_\star)}\right]^2
\left(\frac{\partial l}{\partial r}\right)^{-1}\left[\frac{\partial C_{\rm S} }{\partial r}+2\pi r^3\langle\Sigma\rangle \frac{\partial \Omega_0}{\partial t}\right]
\label{eq:alpha_acc_gen}\\
&= \dfrac{\Omega}{2\pi \Sigma} \left[\dfrac{\mathcal{M}}{v_K(R_\star)}\right]^2
\left(\frac{\partial l}{\partial r}\right)^{-1}\frac{\partial C_{\rm S} }{\partial r}~~~~~~\mbox{in steady state.}
    \label{eq:alpha_acc_steady}
\end{align}

Any local transport mechanism acting as a source of shear viscosity would naturally have $\alpha_{\rm stress}=\alpha_{\rm acc}$, but in our case of the transport being global this is no longer true. We illustrate this difference in Figure~\ref{fig:multi_st}, where we show $\alpha_{\rm stress}$ (middle row) and $\alpha_{\rm acc}$ (bottom row) as a function of time and radial location in the vicinity of the stellar surface for several values of $\M$. This data has also been box-car smoothed in the time dimension only, with a width of $\delta t/2\pi=5$. 

First, one can see that in many cases $\alpha_{\rm stress}$ and $\alpha_{\rm acc}$ are different not only in magnitude but also in sign, in drastic difference with the \citet{SS73} model. In particular, for all $\M$ we find $\alpha_{\rm acc}$ to be positive over extended intervals of time and space, while $\alpha_{\rm stress}$ is negative (clearly visible for $t/2\pi\lesssim 200$, $r<1.4$ in $\M=7$ case, almost always in the $\M=12$ case, and so on). Comparison with the Figures \ref{fig:AM1},\ref{fig:AM2} shows that this situation is typical whenever the parts of the disc adjacent to the star are dominated by the modes trapped in the resonant cavity near the star --- lower and resonant modes --- which carry $C_{\rm S}<0$ and have large $\partial\CS/\partial r>0$. This naturally results in $\alpha_{\rm stress}<0$ and $\alpha_{\rm acc}>0$, see equations (\ref{eq:alpha_stress})-(\ref{eq:alpha_acc_steady}). 

Further from the star, the disc is typically dominated by the upper and vortex-driven modes carrying low-amplitude $C_{\rm S}>0$, which results in $\alpha_{\rm stress}>0$. As for $\alpha_{\rm acc}$, we find it to be negative in some cases, e.g. at $t/2\pi\lesssim 100$ in $\M=12$ case, or at $t/2\pi\sim 100$ in $\M=15$ case, when strong vortex-driven modes are present for $r\gtrsim 1.5$. This situation is typical when $C_{\rm S}>0$ carried by the upper or vortex-driven modes decays sufficiently rapidly with $r$ (as a result of nonlinear dissipation), leading to large $\partial\CS/\partial r<0$ and mass outflow, see (\ref{eq:alpha_acc_gen})-(\ref{eq:alpha_acc_steady}).
But more often we find $\alpha_{\rm acc}>0$ even far from the star, despite $\CS$ being positive there (e.g. for $t/2\pi\lesssim 200$ in $\M=7$ case). This may seem surprising, since $\partial C_{\rm S}/\partial r$ is typically small but negative during such episodes, see Figure \ref{fig:AM2}Be, which would result in mass outflow according to the equation (\ref{eq:alpha_acc_steady}). However, this steady-state equation does not apply in such situations, as the (second) time-dependent term in the brackets in equation (\ref{eq:alpha_acc_gen}) is often more important than the (negative) divergence of $\CS$, driving mass inflow and $\alpha_{\rm acc}>0$ even far from the star.

Second, both $\alpha_{\rm stress}$ and $\alpha_{\rm acc}$ are clearly highly variable both in time and space. This has important implications for the mediation of accretion flow by the modes, which will be discussed in \S \ref{sect:disc_transport}. The peak values of $\alpha_{\rm stress}$ and $\alpha_{\rm acc}$ are reached during the relatively short-lived bursts of activity. Outside of these episodes both $|\alpha_{\rm stress}|$ and $|\alpha_{\rm acc}|$ can maintain much lower levels $\lesssim 10^{-3}$ for extended periods of time.

Third, with our data we can
compare the values of $\alpha$ typical for simulations with different $\M$. One can see that $\alpha_{\rm stress}$ varies roughly from $-10^{-2}$ to $10^{-3}$, with negative (positive) values marking the local dominance of either lower or resonant (upper or vortex-driven) modes. At the same time, $\alpha_{\rm acc}$ only rarely attains negative values (at the level of $-10^{-3}$) implying mass outflow. Its peak amplitude can be as high as $\sim 0.1$ when it is positive. {Results shown in Figure~\ref{fig:multi_st} suggest that $\alpha_{\rm stress}$ and $\alpha_{\rm acc}$ do not show any obvious dependence on $\M$. These transport parameters certainly show little variation with $\M$ at higher $\dot M$, e.g. during the bursts of activity confined to relatively short intervals in the beginning of each simulation. This observation is important since it suggests that similar values of $\alpha_{\rm stress}$ and $\alpha_{\rm acc}$ may be expected also for significantly higher $\M$, typical for many astrophysical objects, see \citetalias{Coleman2021}.}

High amplitudes of $\alpha_{\rm stress}$ and $\alpha_{\rm acc}$ during the initial phases of our simulations lead to rapid re-adjustment of the disc structure in response to the vigorous deposition of the angular momentum carried by the waves, as described in \S \ref{sect:disk_evol}. The top row of Figure~\ref{fig:multi_st} illustrates this process by showing the space-time diagram of the local surface density change in the disc. It is clear that injection of angular momentum by the acoustic modes into the inner disc leads to the depletion of gas in the disc out to $r\sim (1.2-1.6)$, which gets accreted onto the central object. The depletion usually takes place early on, following the large bursts of $\alpha_{\rm acc}$. 

There is also an inverse effect: a significant depletion of mass in the inner disc Figure~\ref{fig:multi_st} acts to suppress the mode amplitudes and to reduce the efficiency of the angular momentum and mass transport. This is quite clear in the $\M=7,9$ simulations, where the amplitudes of $\alpha_{\rm stress}$ and $\alpha_{\rm acc}$ sharply drop right after a substantial gap has been carved up in the inner disc during the initial burst of accretion. In our purely hydrodynamic simulations, there is no mechanism (e.g. MRI) to replenish this accreted material. As a result, we do not achieve a true steady state. However, situation would be different when other transport mechanisms operating in the bulk of the disc are taken into account, see \S \ref{sect:disc_transport}.


\section{Detailed angular momentum balance}
\label{sect:transport_AM_terms}


Next we examine the role played by the different contributions in the angular momentum balance equation (\ref{eq:AM_terms}). In Figure \ref{fig:AM-terms} we plot the time averages (see Appendix \ref{sect:averaging} for details of the averaging procedure) of the three terms featured in that equation, computed for $v_{\phi,0}=v_\Sigma$ (see equation (\ref{eq:v_0-weighted})), for three different simulations ($\M=6,9,12$) listed in the figure. One can see that in all runs the sum of the two terms in the right-hand side of the equation (\ref{eq:AM_terms}), represented by the red dotted curve, falls right on top of the black $\dot M\partial_r l$ curve, as expected. This demonstrates that the relation (\ref{eq:AM_terms}) indeed holds with high accuracy in our simulations. 

\begin{figure}
	\includegraphics[width=\linewidth]{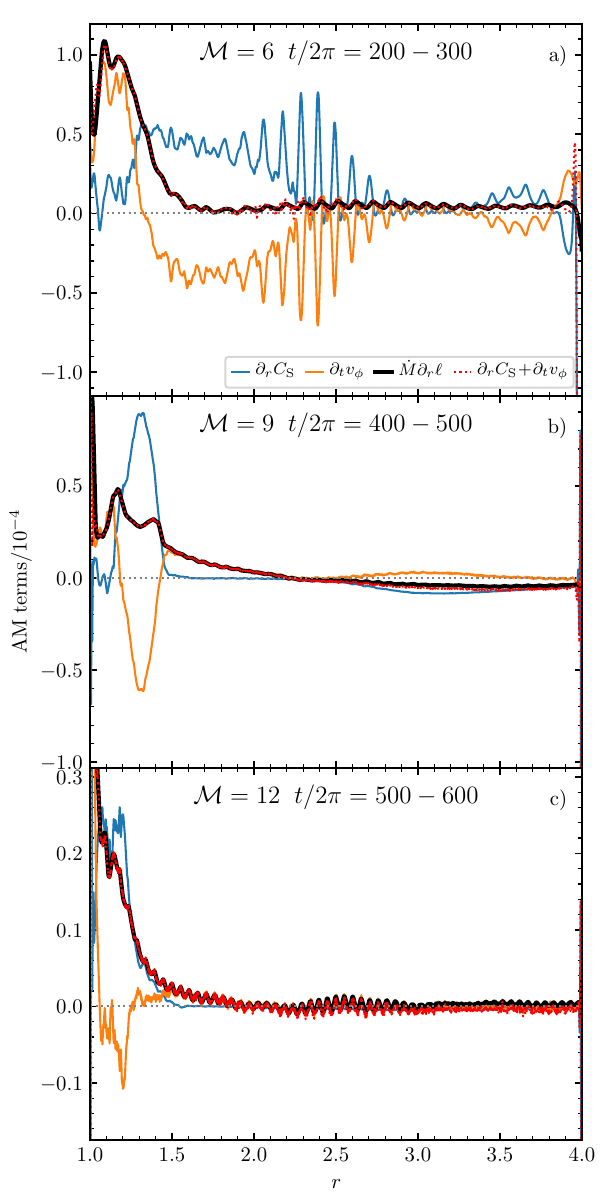}
    \caption{Different terms in the angular momentum balance equation (\ref{eq:AM_terms}) extracted from our simulations. Blue ($\partial_r\CS$), orange (denoted as $\partial_t v_\phi$) and black ($\dot M\partial_rl$) curves correspond to the second, third and first terms in this equation, respectively (see the legend). Red dotted curve gives the sum of the blue and orange curves, which should equal the black one in theory (as it does). Appropriately averaged (see Appendix \ref{sect:averaging} for details) data are shown for 3 runs: (a) $\M=6$ (run M06.HR.r.a) for $t/2\pi=200-300$; (b) $\M=9$ (run M09.FR.r.a) for $t/2\pi=400-500$; (c) $\M=12$ (run M12.FR.mix.a) for $t/2\pi=500-600$. 
    }
    \label{fig:AM-terms}
\end{figure}

In most cases $\dot M$ (as well as $\dot M\partial_r l$, which is plotted) is positive (meaning inflow) and largest near the star, for $r\lesssim 1.5$. This behavior can often be directly related to that of $C_{\rm S}$, e.g. see Figures \ref{fig:AM1}Be, \ref{fig:AM2}Ae,Aj, in which the positive radial derivative of $C_{\rm S}$ provides the dominant contribution to the right hand side of the equation (\ref{eq:AM_terms}) near the star. Large $\partial_r C_{\rm S}>0$ leading to large $\dot M$ arises in the inner disc because of the deposition of the AMF carried by the lower and resonant modes, which are trapped in the resonant cavity near the star. On the other hand, at larger radii one can also find low-amplitude $\dot M<0$, meaning outflow, see Figure \ref{fig:AM-terms}b at $r\gtrsim 2.5$ and \S\ref{sect:alphas}. 

Very importantly, one can see that in all cases shown in Figure \ref{fig:AM-terms} the last term in equation (\ref{eq:AM_terms}) accounting for the time-dependence of the mean rotational profile in the disc (marked $\partial_t v_\phi$ for brevity) plays a very significant role. In many cases its contribution (orange curve) is similar in amplitude (but opposite in sign) to that of $\partial_r C_{\rm S}$, see e.g. Figure \ref{fig:AM-terms}a,b. At first, this may seem rather surprising since the variation of $\Omega$ during the corresponding time intervals is quite small (see Figure \ref{fig:multi_omega}). However, it must be also kept in mind that all terms in equation (\ref{eq:AM_terms}) are {\it second-order} in perturbed variables and are thus small in magnitude. For that reason, even a weak variation of $\Omega$ in time can provide a substantial contribution to the angular momentum balance in equation (\ref{eq:AM_terms}). We will discuss the implications of this observation in \S \ref{sect:disc_transport}.


\section{\texorpdfstring{$C_{\rm S}$-$\dot M$}{CS-Mdot} correlation}
\label{sect:Mdot-CS}


\begin{figure}
	\includegraphics[width=\linewidth]{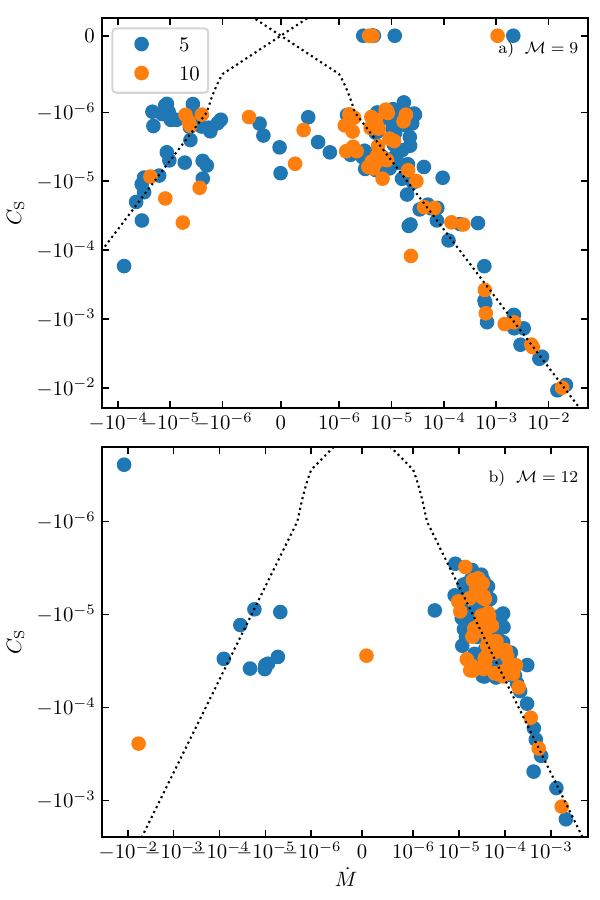}
    \caption{Correlation between the time-averaged values of $C_{\rm S}$ and $\dot M$ for two runs: (a) $\M=9$ run M09.FR.r.a and (b) $\M=12$ run M12.FR.mix.a. Dots show the time-averaged data taken at different moments of time, with colour showing the duration of time-averaging interval: 5 (blue) or 10 (yellow) orbits. The dotted black lines are $C_{\rm S}=\pm 0.5\dot{M}$ to illustrate equation (\ref{eqn:cs_mdot}).
    }
    \label{fig:Mdot-CS}
\end{figure}

A classical steady-state accretion disc with no torque at the origin, which is mediated by the local shear viscosity necessarily exhibits a linear scaling between the {\it instantaneous, local} values of the viscous angular momentum flux $C_{\rm S}$ and $\dot M$ in the form $C_{\rm S}=\dot M l_K$, where $l_K$ is the Keplerian angular momentum. The non-local wave-driven transport that we observe in our BL simulations does not need to obey the same relation, for several reasons. First, the angular momentum balance in equation (\ref{eq:AM_terms}) involves the time-dependent term, which is generally non-zero, see \S\ref{sect:transport_AM_terms}. Second, even in steady state equation (\ref{eq:AM_terms1}) provides only a differential relation between $C_{\rm S}$ and $\dot M$; solving for $\CS$ would require the knowledge of its value at the boundary $r=R_\star$, which is not necessarily zero (as we show next). 

Nevertheless, the overall amplitude of $\dot M$ should still correlate somehow with the amplitude of $C_{\rm S}$, an expectation based on the equation (\ref{eq:AM_terms}). To check whether this is the case we performed the measurement of $C_{\rm S}$ and $\dot M$ at the stellar surface, i.e. $C_{\rm S}(R_\star)$ and $\dot M(R_\star)$. {The radius $r=R_\star$ is chosen so as to minimize the impact of wave damping and non-uniformity of $\dot M$ in the disc on the measurement of angular momentum and mass fluxes.} This measurement is quasi-simultaneous, since we average these quantities over $5$ or $10$ inner orbits to reduce the stochastic noise, which is naturally present near the BL. The results are plotted in Figure \ref{fig:Mdot-CS} for the $\M=9$ and $\M=12$ runs previously discussed in \S\ref{sect:M=9},\ref{sect:M=12}. 

There are several notable features in these plots. First, $C_{\rm S}(R_\star)$ is almost always negative, which is consistent with the $C_{\rm S}(r)$ behavior in Figures \ref{fig:AM1},\ref{fig:AM2}. Second, for moderate negative values of $C_{\rm S}(R_\star)$ one often observes values of $\dot M$, which are roughly equal in magnitude but opposite in sign. This is most likely caused by the oscillatory nature of the wave driven transport activity near the BL (note that Figure \ref{fig:Mdot-CS} uses quasi-logarithmic scale). 

\begin{figure}
	\includegraphics[width=\linewidth]{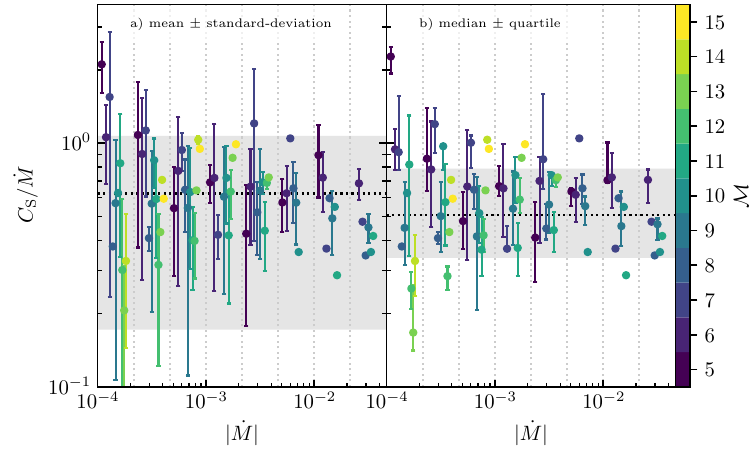}
    \caption{The ratio of $C_{\rm S}/\dot{M}$ at $r=R_\star=1$ (in simulation units, equal to $C_{\rm S}/(\dot{M}\ell_K)|_{r=R_\star}$ in physical units) as a function of $\dot{M}$, with the colour denoting the value of $\mathcal{M}$ of the runs from the data were taken. Each coloured point represents a sample of five orbit period averages of data for the corresponding $\M$, such as blue dots shown in Fig. \ref{fig:Mdot-CS}, and the statistical properties of that sample: in panel (a) we show means $\pm$ one standard-deviation and in (b) we show medians $\pm$ one quartile.
    The data are binned in $\dot{M}$ prior to constructing the sample, the vertical dotted grey lines indicate the bins. Horizontal placement of points within a bin is not meaningful; the points are spread out so that they do not visually overlap. Points without error bars correspond to samples consisting of a single five orbit period average. The horizontal black dotted line and shaded grey band show the global mean/median $\pm$ one standard-deviation/quartile for (a)/(b) respectively. 
    }
    \label{fig:CS_Mdot}
\end{figure}

Third, and most importantly for us, the largest negative values of $C_{\rm S}(R_\star)$ are very tightly correlated with the largest positive values of $\dot M(R_\star)$. We quantified this correlation by examining the dimensionless quantity $C_{\rm S}/\dot{M}$ (in simulation units, with $R_\star=1$, $v_K(R_\star)=1$) in every run of our simulation suite described in \citetalias{Coleman2021}. To evaluate this ratio, we averaged $C_{\rm S}(R_\star)$ and $\dot{M}(R_\star)$ separately in time over five orbits for each simulation\footnote{We also tried using 10 orbits for averaging interval and found that the results were consistent to within $10\%$.}. Part of the reason for this averaging\footnote{We discard some bins to remove the initial data and noisy data in which both quantities frequently change sign.} is to smooth over possible time lags between $C_{\rm S}(R_\star)$ and $\dot{M}(R_\star)$ arising due to the nonlocal nature of the wave driven transport. This data is plotted in Figure \ref{fig:CS_Mdot}, and results in (median $\pm$ one quartile) 
\begin{align}
\label{eqn:cs_mdot}
    \left.\dfrac{C_{\rm S}}{ \dot{M}\ell_K}\right|_{r=R_\star}=- 0.51^{+0.28}_{-0.17}.
\end{align}
This relation (shown with dotted lines in Figures \ref{fig:Mdot-CS},\ref{fig:CS_Mdot}) once again demonstrates that mass inflow into the BL requires {\it negative} $C_{\rm S}$ at the stellar interface. It also characterizes the efficiency with which mass accretion is enabled by the angular momentum injection into the disc by the wave modes excited in the BL. Finally, it provides an important boundary condition, which can be used in constructing semi-analytical models describing the structure of the inner regions of accretion discs fully or partly mediated by the acoustic waves.

\section{Wave-driven disc evolution}
\label{sect:disk_evol}


\begin{figure*}
	\includegraphics[width=\linewidth]{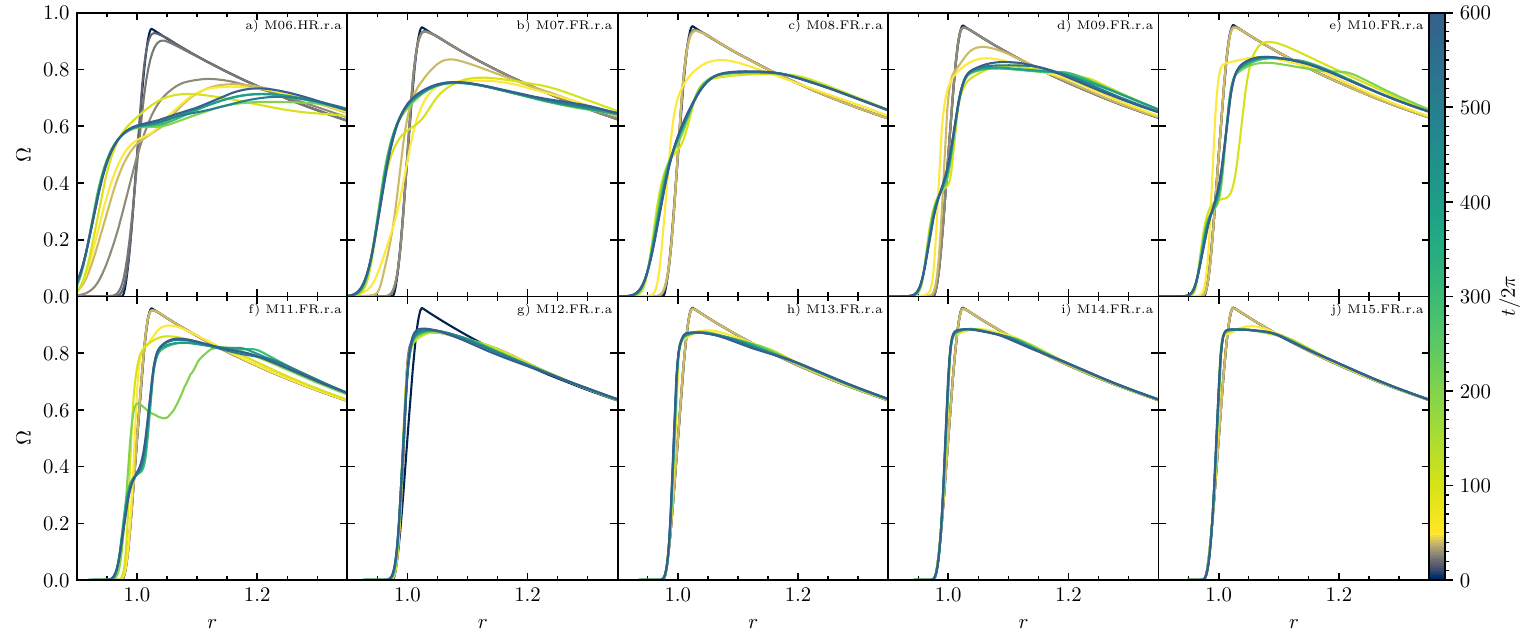}
    \caption{Evolution of $\Omega(r)$ profiles in simulations with different values of $M$ (found in the run label). Different curves correspond to different moments of time coded in the colourbar on the right. See text for details.}
    \label{fig:multi_omega}
\end{figure*}
\begin{figure}
	\includegraphics[width=\linewidth]{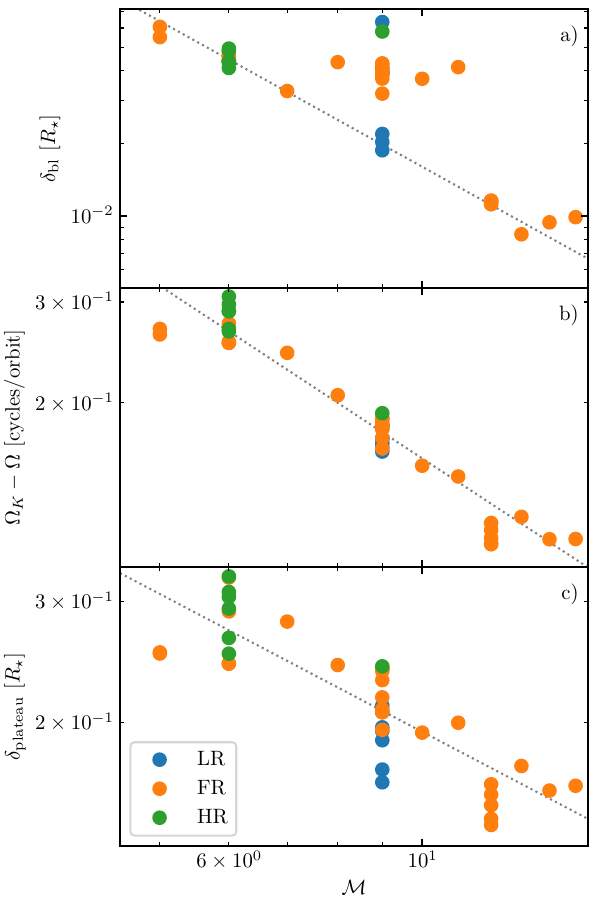}
    \caption{(a) Width of the BL as a function of $\M$; dotted line corresponds to equation (\ref{eq:deltaBL}). 
    (b) Deviation of the plateau from $\Omega_K(R_\star)$ as a function of $\M$; dotted line corresponds to equation (\ref{eq:deltaOmega}). 
    (c) Radial extent of the $\Omega$ plateau as a function of $\M$; dotted line corresponds to equation (\ref{eq:deltaPlateau}). The dotted lines are by-eye fits to the relevant data.
    The data shown correspond to the final output file of each simulation ($t/2\pi=600$).}
    \label{fig:BLproperties}
\end{figure}

In the absence of wave damping $C_{\rm S}$ would be exactly conserved in our globally isothermal setup, implying no transfer of the wave angular momentum to the disc and $\dot M=0$. But in our simulations wave modes of different types can exchange their angular momentum with the disc fluid, causing its surface density to evolve. Since our runs have no viscous or radiative dissipation, this exchange is accomplished through the nonlinear dissipation of the waves \citep{GR01,R02}: nonlinear steepening of the acoustic wave profile eventually turns the wave into a shock \citep{LL}, transferring its angular momentum to the background flow \citep{RRR16}. This process inevitably occurs even if the shock is weak. 

Nonlinear damping always {\it reduces} wave amplitude and $|C_{\rm S}|$, regardless of the sign of $C_{\rm S}$. As a result, in agreement with the equations (\ref{eq:AM_terms})-(\ref{eq:AM_terms1}), non-zero $\dot M$ arises in the disc changing its structure. But the direction of the mass flow also depends on a particular type of the wave and, more specifically, on the sign of AMF $C_{\rm S}$ that it carries.  

As we saw in \S\ref{sect:alphas}, damping of the modes with $C_{\rm S}<0$, e.g. lower and resonant modes (see \S\ref{sect:transport_AM_modes}), gives rise to $\partial C_{\rm S}/\partial l\propto \partial C_{\rm S}/\partial r>0$, leading to $\dot M>0$ (see equations (\ref{eq:AM_terms}),(\ref{eq:AM_terms1})), i.e. mass {\it inflow} through the disc. On the other hand, dissipation of the modes with $C_{\rm S}>0$, i.e. the upper and vortex-driven modes, can sometimes lead to $\partial C_{\rm S}/\partial l>0$. As a result, the disc fluid gains angular momentum, resulting in $\dot M<0$, i.e. mass {\it outflow}. 

In practice, lower and resonant modes carrying $C_{\rm S}<0$ generally affect disc evolution much stronger than the other types of modes. Since these modes are trapped in the resonant cavity in the inner disc, gas accretion onto the star driven by them clears out a substantial depression, or gap, in the inner disc. The development of such a gap is illustrated in the top row of Figure~\ref{fig:multi_st}. One can see that the reduction of $\Sigma$ can be quite substantial, with the gap depth reaching $\sim 80\%$ of the original density at $r\lesssim 1.5$ in simulations with $\M\lesssim 9$. For higher-$\M$ runs, which often exhibit more regular patterns of the acoustic modes, the decrement of $\Sigma$ is not so large, with the gap depths at the level of $\sim (20-30)\%$ being more typical, see Figure~\ref{fig:multi_st}. The radial extent of the gap also decreases with increasing $\M$, in agreement with the smaller radial width of the resonant cavity at higher $\M$, see \S\ref{sect:transport_AM_modes}.

In our simulations the deepening of the gap saturates once it reaches a sufficient depth that depends on $\M$. The gap does not get refilled by the matter arriving from larger radii, as would be expected in a real accretion disc, because our runs do not have an an explicit mechanism of the angular momentum transport in the outer disc (i.e. $\alpha$-viscosity or MRI). In real discs one should expect the gap to be shallower than what we find in our simulations.


\subsection{Evolution of the \texorpdfstring{$\Omega(r)$}{Omega} profile}
\label{sect:omega_evol}


Sharp gradients of $\Sigma$ that develop around the gap region cause substantial modification of $\Omega(r)$ profile away from the purely Keplerian $\Omega_K(r)$ in the bulk of the disc. In a steady state $\Omega(r)$ is given by
\begin{align}
\Omega^2(r)=\Omega_K^2(r)+\frac{1}{\Sigma r}\frac{\partial P}{\partial r},
\label{eq:Omega}
\end{align}
following from the radial momentum balance equation. Because of the initial non-uniform profile of $\Sigma$ and the resultant radial pressure gradient, some deviation of $\Omega$ from $\Omega_K$ is present even at the start of simulations. 

But after the gap develops in the disc, these $\Omega$ deviations increase as the second term in the right hand side of the equation (\ref{eq:Omega}) becomes much larger. This is illustrated in Figure \ref{fig:multi_omega}, which shows the profiles of $\Omega(r)$ at different moments of time in simulations with different $\M$. Using the simulation data on $\Sigma(r)$ (e.g. the ones shown in the top row of Figure \ref{fig:multi_st}) we find that the $\Omega(r)$ profile agrees very well with the equation (\ref{eq:Omega}). Interestingly, equation (\ref{eq:Omega}) describes the $\Omega(r)$ behavior quite well even inside the BL, where $\Omega(r)$ substantially deviates from $\Omega_K$. This means that even within the BL the radial velocity of the accreting matter $v_r$ is still quite small, so that the inertial terms in the radial momentum balance equation can be neglected.

In agreement with the equation (\ref{eq:Omega}), we find $\Omega<\Omega_K$ interior from the deepest part of the gap, where $\Sigma$ and $P$ decrease with radius. In fact, $\Omega$ tends to exhibit a relatively flat, {\it plateau}-like segment with $\Omega(r)\approx \Omega_{\rm max}$ just outside the BL. 
Right outside the deepest part of the gap, $P$ increases with $r$ (very strongly for lower $\M$), driving $\Omega$ {\it above} $\Omega_K$. This can be most easily seen in Figure \ref{fig:multi_omega}c-f. As the gap width and depth become smaller for higher $\M$, so does the deviation of $\Omega$ from $\Omega_K$: the region near the BL where $\Omega(r)$ exhibits a plateau gets narrower as $\M$ grows, and the value of $\Omega$ in this region gets closer to $\Omega_K$. 

To characterize these $\Omega(r)$ features developing near the BL, in Figure \ref{fig:BLproperties} we plot the width of the BL $\delta_{\rm BL}$, defined as the radial extent over which $\Omega$ transitions from 25\% to 75\% of its maximum value, i.e. $\delta_{\rm BL}\equiv R\left(\Omega=0.75\Omega_{\rm max}\right) - R\left(\Omega = 0.25\Omega_{\rm max}\right)$
(panel (a)); the deviation of $\Omega(r)$ in the plateau region $\delta\Omega$ from $\Omega_K(R_\star)=1$, i.e. $\delta\Omega\equiv 1-\Omega_{\rm max}$ (panel (b)); and the radial width $\delta_{\rm plateau}$ of the plateau in $\Omega$, defined as the radial extent over which $\Omega\ge 0.9\Omega_{\rm max}$. One can see that all three plotted variables generally {\it decrease} with increasing $\M$, as mentioned previously. Note that some values of $\M$ for which we have multiple simulations show a spread in these derived variables, see e.g. the variation of $\delta_{\rm BL}$ for $\M=9$ in panel (a). 

Such spreads in $\delta_{\rm BL}$ are caused by the emergence of an inflection point in $\Omega(r)$ profile inside the BL in some runs, which acts to increase the radial range over which $\Omega$ varies. This feature is clearly seen e.g. in Figure \ref{fig:multi_omega}d-f (see \citealt{BRS12} {and \citealt{Dittmann2021}} for similar observations). To circumvent the effect of this feature on the determination of $\delta_{\rm BL}$ we focus on the smallest values of $\delta_{\rm BL}$ for a given $\M$. To guide the eye, in Figure \ref{fig:BLproperties}a we run\footnote{This and other dotted lines in Figure \ref{fig:BLproperties} are not formal fits to the data.} a dotted line 
\ba
\delta_{\rm BL}\approx 2~\M^{-2}
\label{eq:deltaBL}
\ea  
through these points that appears to provide a decent fit to the lower envelope of the measured $\delta_{\rm BL}$ values. As previously noted in \citet{BRS12}, this scaling implies that $\delta_{\rm BL}$ is equal to a fixed (i.e. independent of $\M$) number of {\it stellar} scaleheights (equal to $\M^{-2}$ in our notation). In our case this number is $\approx 2$, which is significantly less than $7-8$ found in \citet{BRS12}, a difference naturally explained by the fact that \citet{BRS12} did not attempt to account for the effect of the inflection point on the $\Omega$ profile. 

In Figure \ref{fig:BLproperties}b we plot $\delta\Omega$ --- the deviation of $\Omega$ from unity in the plateau region --- together with a by-eye fit 
\ba
\delta\Omega\approx 1.5~\M^{-1}. 
\label{eq:deltaOmega}
\ea  
This rough scaling can be understood analytically using equation (\ref{eq:Omega}) if we also assume that the radial scale of $\Sigma$ variation in the gap near the stellar surface is of order of the {\it disc} scaleheight $\M^{-1}$, which appears to be the case in our runs. 
The data for the radial extent of $\Omega$ plateau $\delta_\mathrm{plateau}$ is shown in Figure \ref{fig:BLproperties}c together with a simple fit (dotted curve)
\ba  
\delta_\mathrm{plateau}\approx 0.9~\M^{-2/3}.
\label{eq:deltaPlateau}
\ea  
Based on the Keplerian rotation law and equation (\ref{eq:deltaOmega}), one might have expected a steeper dependence on $\M$ for a flat $\Omega$ plateau, closer to $\propto\M^{-1}$. However, examination of Figure \ref{fig:multi_omega} shows that approximating the  $\Omega(r)$ plateau shape as flat is very approximate, and that it has a certain non-trivial radial dependence to it. Thus, the exponent in equation (\ref{eq:deltaPlateau}) different from $-1$ should not come as a surprise. 

{Knowledge of $\delta\Omega$ and $\delta_\mathrm{plateau}$ is important since we often find the plateau region to contain near-surface vortices that drive prominent spiral arms in the disc, see \citetalias{Coleman2021}. The pattern speed of such features should be roughly equal to $1-\delta\Omega$ (as the vortices are comoving with the fluid), while $\delta_\mathrm{plateau}$ informs us about the distance out to which these near-surface vortices might be expected to be found.}

Given that the exact shape of the $\Omega$ plateau depends on the $\Sigma(r)$ profile in the gap, one may wonder if scalings (\ref{eq:deltaOmega}), (\ref{eq:deltaPlateau}) would still hold in the presence of mass inflow from larger radii, which is expected in real discs. It is very likely that $\delta\Omega$ and $\delta_{\rm plateau}$ behaviors shown in Figure \ref{fig:BLproperties} will change in that situation. However, the scaling (\ref{eq:deltaBL}) for the width of the BL ($\delta_{\rm BL}$) must be robust regardless of the presence or absence of the mass inflow, since the existence of the BL is independent of the presence or absence of a depression in $\Sigma$ outside the star.


\section{Discussion}
\label{sect:disc}


Results of the previous sections illustrate the important effect of the wave modes triggered by the supersonic shear inside the BL on the angular momentum and mass transport in the inner disc. We now discuss some additional aspects of the wave-driven transport processes in the vicinity of the BL.


\subsection{Modes dominating mass transport and disc evolution}
\label{sect:disc_transport}


The discussion in \S\ref{sect:transport-lower}-\ref{sect:transport-other} allows us to identify the modes most relevant for driving the accretion onto the star. Since accretion implies $\dot M>0$, we can immediately conclude that it is the lower acoustic and resonant modes that must be responsible for driving the mass inflow onto the stellar surface. This conclusion is supported by Figure \ref{fig:Mdot-CS} and the results presented in \S\ref{sect:Mdot-CS}, which show that the highest values of $\dot M$ require $\CS<0$ , which is a unique feature of these two types of modes.   

At the same time, the vortex-driven (and, at lower amplitude, the upper acoustic) modes drive a weak mass outflow in the disc, which is not conducive to accretion. In realistic discs this outflow would be suppressed by the mass inflow caused by other sources of the angular momentum transport capable of operating in the bulk of the disc, e.g. the MRI. 

It is important that the $\dot M$ driven by the lower acoustic and resonant modes is large only in radially narrow region of the disc near the stellar surface, \newedit{ where these modes are trapped. This means that these modes by themselves are not capable of maintaining constant $\dot M$ through the whole disc, which would be necessary to keep it in a steady state. Other transport mechanisms (e.g. the MRI) are clearly needed to ensure a steady delivery of mass through the disc towards the BL with the  radially-constant $\dot M$. }

It is also important to realize that in realistic discs, inside the resonant cavity where the $\dot M$-driving modes are trapped, the angular momentum transport would be provided {\it both} by the waves, globally, {\it and} by the MRI, locally (since $d\Omega/dr<0$ in that region, see Figure \ref{fig:multi_omega}). This enhanced transport is likely to result in a non-trivial structure of the surface density in this near-BL part of the disc. \newedit{ Moreover, MRI turbulence can serve as a driver of additional waves in the disc \citep{Heinemann2009}, which might interact with the acoustic modes excited in the BL.}


\subsection{Global nature of the angular momentum transport}
\label{sect:global}


Our morphological study in \citetalias{Coleman2021} demonstrates that the waves excited in the vicinity of the BL can propagate over substantial distances in the disc. This is clear for the vortex-driven and upper acoustic modes that can propagate from their corotation region near the BL all the way into the outer disc where they get gradually damped. However, even the modes which are trapped near the BL --- the lower acoustic waves and the resonant modes --- still travel over a significant radial range in the disc, comparable to $R_\star$. This is especially true for the low-$m$ resonant modes, for which the corotation radius can lie quite far out in the disc, see \citetalias{Coleman2021}. 

The long-range propagation of the waves makes the angular momentum transport effected by them truly global. 
It gives rise to a non-local coupling between the BL, where the accreting matter uploads its angular momentum and energy to the waves, and the more distant parts of the disc, where the waves dissipate, driving $\dot M$ and injecting their energy into the disc fluid \citep{RRR16}. Such non-locality is typical also for the angular momentum transport by the planet-driven density waves in the protoplanetary discs \citep{Lunine1982,GR01,R02,RP2012,PR2012}.

Because of its non-locality, the wave-driven transport cannot be characterized by a single $\alpha$ parameter, which is often invoked to describe local transport \citep{SS73,BalPap1999}. As we demonstrate in \S\ref{sect:alphas}, the values of $\alpha$ near the BL computed using stress $\CS$ and $\dot M$ (which are the same for local transport) are different not only in magnitude but also in sign, see Figure \ref{fig:multi_st}. For that reason, the standard methods for calculating the structure and viscous evolution of the disc based on the diffusion-type equation approach \citep{Lynden1974}, would fail to describe the disc in the vicinity of the BL. Instead, calculations of the wave-mediated structure of the inner disc need to explicitly account for the long-range propagation and dissipation of the waves excited in the BL.


\subsection{Observational implications}
\label{sect:observations}


Non-local wave-driven transport effected by the waves has important implications for the observational appearance of the objects accreting through the BL. Models of the BL structure using the local  $\alpha$-viscosity inevitably predict that local energy dissipation in the layer should heat gas in the BL to high temperatures \citep{POP93,NAR93}. This would naturally give rise to a high-energy component in the spectra of objects accreting through the BL, which is often not observed \citep{Ferland1982}.  

Energy transport by the waves changes the whole pattern of the energy dissipation: it occurs not in the narrow BL itself, but over a larger area of the disc, lowering the effective temperature of emission associated with the BL and naturally alleviating the "missing boundary layer" problem \citep{Ferland1982}. The rate $d\dot E/dr$ (per unit time and radius) at which thermal energy is deposited in the disc by a wave (with pattern speed $\Omega_P$) is closely related to the rate of dissipation of the wave angular momentum flux, as $d\dot E/dr=(\Omega-\Omega_P)d\CS/dr$ \citep{Lynden1972,GR01,RRR16}. Thus, the results of our study on the radial profiles of $\CS$ for different wave modes can be directly employed to analyze the spatial distribution of the disc heating induced by the BL-excited waves. 


\subsection{Comparison with the existing studies}
\label{sect:literature}


In the past, \citet{BRS12,BRS13a,BRS13b} analyzed the behavior of the wave angular momentum flux $\CS$ in the vicinity of the BL, and related it to the $\dot M$ behavior and the wave amplitude in the disc. {More recently, \citet{Dittmann2021} looked at the $\CS$ behavior in the disc (as well as the flow of mass and angular momentum into the star) for non-zero stellar spin rates.} Our present work goes beyond these studies in several ways. 

First, we analyze $\CS$ behavior for a variety of {\it individual} modes present in the disc. Coupling this with the results of our morphological study in \citetalias{Coleman2021} allows us to understand the $\CS$ behavior for each mode type --- lower and upper, resonant, vortex-driven, and so on. Second, we follow how the behavior of $\CS$ changes as our simulations progress and different types of wave modes come and go. Third, we systematically explore the differences in the $\CS$ behavior as a function of the Mach number $\M$ of our runs.


\section{Summary}
\label{sect:sum}


In this work we studied the angular momentum and mass transport driven by the different types of the waves operating in the vicinity of the BL of an accretion disc. Our analysis is based on a large suite of global, 2D hydrodynamic simulations described in \citetalias{Coleman2021}, in which a large number of modes have been previously identified and characterized. Results of our present study can be summarized as follows.

\begin{itemize}

\item In wave-mediated inner region of the disc mass accretion rate $\dot M$ has a significant contribution arising from the correlated radial velocity $v_r$ and the surface density perturbation $\delta\Sigma$. 

\item {The efficiency of angular momentum transport expressed through the effective $\alpha$-parameter appears to not depend strongly on the Mach number of the flow $\M$.}

\item By examining angular momentum flux $C_{{\rm S},m}$ carried by each individual Fourier harmonics, we were able to determine the transport properties associated with each type of the mode for different values of the Mach number $\M$. In particular, we find that lower acoustic and resonant modes carry negative angular momentum flux, whereas the vortex-driven and upper acoustic modes carry positive angular momentum flux.

\item Nonlinear damping of the modes leads to mass inflow ($\dot M>0$) for the lower acoustic and resonant modes, and weak mass outflow for the vortex-driven and upper acoustic modes. This implies that accretion onto the central object must be mediated by a combination of the lower acoustic and resonant modes.

\item Wave-driven transport is intrinsically non-local, which leads to the values of effective $\alpha$-parameter determined through stress and $\dot M$ being different not only in amplitude but also in sign. This is very different from the conventional local transport for which these values are the same.

\item Despite the non-local nature of the wave-mediated transport we still find a strong correlation, given by the equation (\ref{eqn:cs_mdot}), between the angular momentum flux injected at the BL into the disc, and the mass accretion rate through the BL.
    
\item Despite the long duration of our simulations, we find the time-dependent contribution to the angular momentum balance caused by the variation of $\Omega$ (see equation (\ref{eq:AM_terms})) to play an important role (see \S\ref{sect:transport_AM_terms}).    

\item We characterize the wave-driven evolution of the disc properties --- surface density $\Sigma$ and angular frequency $\Omega$ profile --- as a function of time and $\M$. 

\end{itemize}

Our study, based on 2D hydro simulations, provides a natural foundation for understanding the properties of the wave-driven transport in future three-dimensional and fully MHD simulations of the BLs. 

\section*{Acknowledgements}

We thank Jim Stone for making the \athena code publicly available. We gratefully acknowledge financial support from NSF via grant AST-1515763, NASA via grant 14-ATP14-0059, and Institute for Advanced Study via the John N. Bahcall Fellowship to R.R.R. Research at the Flatiron Institute is supported by the Simons Foundation. Resources supporting this work were provided by the NASA High-End Computing (HEC) Program through the NASA Advanced Supercomputing (NAS) Division at Ames Research Center. Through allocation AST160008, this work used the Extreme Science and Engineering Discovery Environment (XSEDE), which is supported by National Science Foundation grant number ACI-1548562 \citep{XSEDE}.

\section*{Data Availability}

The data underlying this article will be shared on reasonable request to the corresponding author.




\bibliographystyle{mnras}
\bibliography{citations}



\appendix


\section{Details of the time-averaging procedures}
\label{sect:averaging}


Here we summarize the technical details of various averaging and smoothing procedures that were used in analyzing the data and producing various plots. 
For all the line plots in this paper (except for Fig.  \ref{fig:multi_omega}) we used the unbinned FFT data to measure azimuthal averages ($m=0$; all the quantities inside angular brackets) and higher order $m>0$ modes with high cadence (exactly 20 outputs per orbit with non uniform cadence; see \citetalias{Coleman2021} for more details). As the final step of the calculation we average in time over the intervals denoted in the figures.
For Figs.~\ref{fig:AM1} and \ref{fig:AM2}
we chose to plot only the five modes with the largest time (over the interval noted in the figure) and radially (over $r>1$) integrated $|C_{{\rm s},m}|$.

We examine Eqn.~(\ref{eq:AM_terms}) in Fig.~\ref{fig:AM-terms}. We rewrite and expand this equation below to be explicit about the averaging done in the figure: 
\begin{align}
\nonumber
\underbrace{2\pi r \left\langle\Sigma v_r\right\rangle \pder{}{r}\left(r \dfrac{\left\langle \Sigma v_\phi \right\rangle}{\left\langle \Sigma \right\rangle} \right)}_{\dot{M}\partial_r\ell}&= 
\underbrace{\frac{\partial C_{\rm S} }{\partial r}}_{\partial_r C_{\rm S}}+
\underbrace{2 \pi r^2\langle\Sigma\rangle \pder{}{t}\dfrac{\left\langle \Sigma v_\phi \right\rangle}{\left\langle \Sigma \right\rangle}}_{``\partial_t v_\phi"},
\end{align}
where $C_{\rm S}$ is {computed} using Eqns.~(\ref{eq:C_S}) and (\ref{eq:v_0-weighted}), and the underbraces are used to denote how the terms are labeled in Fig.~\ref{fig:AM-terms}. All the azimuthally averaged quantities (i.e. the ones in angular brackets) are measure directly from the $m=0$ mode of our high cadence FFT data as explained above. The partial derivative with respect to time is computed via finite differences on the high cadence FFT data to second order accuracy. After these computations are completed the data is averaged in time over the time interval shown in the figure.

\bsp	
\label{lastpage}
\end{document}